\documentclass[iop,apj,numberedappendix,appendixfloats,twocolappendix]{emulateapj}
\usepackage{mathtools}
\usepackage{graphicx}
\usepackage{physics}
\usepackage{xcolor}
\usepackage{hyperref}

\begin{document}

\shorttitle{Helium Reionization Simulations I}
\shortauthors{La Plante \& Trac}

\title{Helium Reionization Simulations. I. Modeling Quasars as Radiation Sources}
\author{Paul La Plante\altaffilmark{*} and Hy Trac}
\affil{McWilliams Center for Cosmology, Department of Physics, Carnegie Mellon University, Pittsburgh, PA 15213, USA}
\altaffiltext{*}{plaplant@andrew.cmu.edu}



\begin{abstract}
  We introduce a new project to understand helium reionization using fully
  coupled $N$-body, hydrodynamics, and radiative transfer simulations. This
  project aims to capture correctly the thermal history of the intergalactic
  medium (IGM) as a result of reionization and make predictions about the
  Lyman-$\alpha$ forest and baryon temperature--density relation. The dominant
  sources of radiation for this transition are quasars, so modeling the source
  population accurately is very important for making reliable predictions. In
  this first paper, we present a new method for populating dark matter halos
  with quasars. Our set of quasar models includes two different light curves, a
  lightbulb (simple on/off) and symmetric exponential model, and
  luminosity-dependent quasar lifetimes. Our method self-consistently reproduces
  an input quasar luminosity function given a halo catalog from an $N$-body
  simulation, and propagates quasars through the merger history of halo
  hosts. After calibrating quasar clustering using measurements from the Baryon
  Oscillation Spectroscopic Survey, we find that the characteristic mass of
  quasar hosts is $M_h \sim 2.5 \times 10^{12}\ h^{-1}M_\odot$ for the lightbulb
  model, and $M_h \sim 2.3 \times 10^{12}\ h^{-1}M_\odot$ for the exponential
  model. In the latter model, the peak quasar luminosity for a given halo mass
  is larger than that in the former model, typically by a factor of 1.5--2. The
  effective lifetime for quasars in the lightbulb model is 59 Myr, and in the
  exponential case, the effective time constant is about 15 Myr. We include
  semi-analytic calculations of helium reionization, and discuss how to include
  these quasars as sources of ionizing radiation for full hydrodynamics with
  radiative transfer simulations in order to study helium reionization.
\end{abstract}


\section{Introduction}
\addtocounter{footnote}{-1} 
\addtocounter{Hfootnote}{-1}
\label{sec:intro}
Helium reionization is an important epoch in the Universe's history, and the
most recent large-scale transition of the intergalactic medium (IGM). During the
epoch of hydrogen reionization, the first stars and galaxies emitted photons
capable of ionizing hydrogen and singly ionizing helium (whose ionization
energies are 13.6 and 24.6 eV, respectively). However, the spectra of these
first sources did not contain a sufficient number of high-energy photons capable
of doubly ionizing helium, which requires a much larger ionization energy (54.4
eV). Consequently, helium was predominantly singly ionized following hydrogen
reionization until a burst of quasar activity at redshifts
$6 \gtrsim z \gtrsim 2$. Quasars are thought to be the first objects to emit an
appreciable number of photons capable of doubly ionizing helium. However,
because the birth of quasars requires additional time for structure to form and
sufficient mass to assemble inside dark matter halos, this period of evolution
occurs later in the Universe's history.

Recent and upcoming efforts to look for quasars include the Baryon Oscillation
Spectroscopic Survey (BOSS) of SDSS-III \citep{dawson_etal2013}, the Hyper
Suprime Cam of the Subaru telescope \citep{kashikawa_etal2015}, and DESI
\citep{schlegel_etal2011}. There are currently about 420,000 unique quasar
objects \citep{flesch2015}, with this number projected to increase by an order
of magnitude after the conclusion of the next generation of experiments. This
rich set of observations allows us to characterize quasars to an unprecedented
level of accuracy, and better characterize their properties. This is especially
true at high redshift ($z \gtrsim 6$), where there are currently few
observations. Determining quasar properties at high redshifts is helpful for
understanding the growth of structure, as well as providing observations of
reionization through measuring their absorption spectra.

Observations have shown that quasar activity peaks between
$2 \lesssim z \lesssim 3$ \citep{warren_etal1994,schmidt_etal1995}. The
Gunn--Peterson trough \citep{gunn_peterson1965} of helium has been detected at
$z > 3$ \citep{jakobsen_etal1994,zheng_etal2008,syphers_shull2014}, implying
that some fraction of helium was still present as He \textsc{ii} at these
redshifts. Helium absorption then transitions to becoming patchy, with extended
regions of absorption and transmission in the He \textsc{ii} Lyman-$\alpha$
forest \citep{reimers_etal1997}, and seems to be completed by $z \sim 2.7$
\citep{dixon_furlanetto2009,worseck_etal2011}, which coincides with the peak in
quasar activity. However, to observe the Gunn--Peterson trough of He \textsc{ii},
the sight line must be free of any intervening Lyman-limit systems. This means
that the number of observations for these measurements is rather small (of
$\order{10}$).

When discussing helium reionization, it is important to understand the
properties of the ionization sources, such as quasars' lifetimes and light
curves. On the theoretical side of the problem, there are some predictions for
quasar properties, but also a fair degree of uncertainty. By treating quasars as
accretion disks around super-massive black holes (SMBHs), one can show that the
maximal conversion efficiency $\epsilon$ for converting mass to luminosity is
$\epsilon \sim 0.3$ \citep{thorne1974}. Further, for matter accreting onto an
SMBH at the Eddington limit \citep{eddington1926}, one obtains an exponential
increase in mass and luminosity with a characteristic time scale (called the
Salpeter $e$-folding time) of $\tau = 45$ Myr for $\epsilon=0.1$
\citep{salpeter1964,wyithe_loeb2003b}. Cosmological simulations that seek to
capture the relationship between quasars and their galaxy hosts have treated
quasar activity as being the result of a major-merger event between two galaxies
\citep{springel_etal2005,hopkins_etal2006,hopkins_etal2008}, or a cold-flow
accretion of gas onto the central SMBH \citep{dimatteo_etal2012}. However, there
is no definitive evidence that quasars accrete exclusively at the Eddington
limit, or are limited to a single episode of highly luminous activity.

Observations can also help us understand the physics of quasars, though
typically at larger scales than theory or simulation. Since the entire rise and
fall of quasar number density spans a time of roughly 10$^9$ years, the quasar
lifetime must be shorter than this \citep[][p. 324]{osmer2004}. At the other
extreme, observations of the quasar proximity zone show that quasar lifetimes
should be at least 10$^5$ years \citep[][p. 169]{martini2004}. This time scale
corresponds to the photoionization timescale of relatively high-density neutral
hydrogen systems observed to be ionized in the IGM, and so the lifetime of the
quasar must be at least this long in order to maintain the highly ionized level
of these systems observed in the Lyman-$\alpha$ forest. Further constraints are
difficult to obtain, and usually rely on indirect methods such as quasar
clustering measurements (\textit{e.g.},
\citealt{porciani_etal2004,porciani_norberg2006,white_etal2012}). Estimates made
using these methods yield values for the quasar lifetime that are 10--100 Myr,
with most values being $\sim$30 Myr, which is comparable to the Salpeter
$e$-folding time. Further, there are few definitive constraints on quasar light
curves (though see \citealt{hopkins_hernquist2009}).

For the universal populations of quasars, the major pieces of data are their
number density as a function of luminosity and redshift (\textit{i.e.}, the
quasar luminosity function (QLF) $\phi(L,z)$, \textit{e.g.},
\citealt{schmidt_green1983,boyle_etal2000,ross_etal2013}), and their spatial
clustering \citep{outram_etal2003,porciani_etal2004,white_etal2012}. These
observations can constrain scaling relations between quasars and their hosts
(\textit{e.g.}, \citealt{conroy_white2013}), or used to calibrate subgrid models
for simulations (\textit{e.g.}, \citealt{feng_etal2014}). However, as mentioned
above, the properties of individual quasars are difficult to extract from these
observations, due to degeneracies. The imposed constraints are typically weak,
and only provide order-of-magnitude precision.

Cosmological simulations are an ideal tool for furthering our knowledge about
this portion of the universe's history. Helium reionization leaves a lasting
impression on the thermal history of the IGM: the relative hardness of quasar
spectra means that there is a large degree of photoheating of the IGM while
reionization is occurring. Thus, it is important to include hydrodynamics in
simulations, in order to include the effects of baryonic physics. Additionally,
due to the relatively long mean free path of far-UV and soft X-ray photons when
looking at helium reionization, it becomes important to include radiative
transfer calculations in simulations. Thus, semi-analytic calculations that
assume a sharp reionization front are typically poor approximations of the
physical situation. Even 1D radiative transfer codes are not realistic enough to
calculate the inhomogeneous reionization process, especially when reionized
regions begin to overlap. Due to the highly biased nature of quasar sources,
this is typically early in the reionization process. Therefore, 3D radiative
transfer calculations are essential for capturing the complicated physics of
helium reionization. As mentioned earlier, the large degree of thermal heating
argues for simulations in which the hydrodynamics calculations are coupled to
the radiative transfer ones. This work builds on and extends previous
investigations of helium reionization, which either were semi-numerical
\citep{furlanetto_oh2008,dixon_etal2014} or applied radiative transfer in
post-processing
\citep{mcquinn_etal2009,mcquinn_etal2011,compostella_etal2013,compostella_etal2014}.

Our approach to helium reionization uses simulations, with $N$-body,
hydrodynamics, and radiative transfer solved simultaneously. An essential first
step of this calculation is to understand the sources of reionization, and
ensure that their properties match the observations as nearly as possible. To
this end, we use the observed QLF from the SDSS and the COSMOS survey across
various redshift epochs
(\citealt{masters_etal2012,mcgreer_etal2013,ross_etal2013}, hereafter M12, M13,
and R13) and the clustering measurements from BOSS \citep{white_etal2012} to
inform the properties of individual quasars for our simulation input. By using
these two constraints, as well as a formalism for populating dark matter halos
with quasars that we will outline below, we are able to select simulated quasar
hosts that agree well with the latest observational constraints. Specifically,
matching the QLF means that we have an observationally accurate number of
ionization sources, and matching the clustering measurements means our topology
of reionization (\textit{e.g.}, the size and overlap of reionized regions) will
be similar to the actual reionization process. The clustering can also have an
effect on the spatial correlations present in the radiation field, which can
affect the baryon acoustic oscillation (BAO) measurement from the Lyman-$\alpha$
forest.

This first paper of the series discusses the way in which we create sources for
our simulations of helium reionization. In Sec.~\ref{sec:pop}, we describe our
simulation strategy, and how we construct a quasar catalog from an $N$-body halo
catalog. In Sec.~\ref{sec:clustering}, we explain how we modify our quasar
properties in order to match recent observations. In Sec.~\ref{sec:discussion},
we explore implications of our findings for quasar populations. In
Sec.~\ref{sec:reion}, we discuss implications for helium reionization. Finally,
in Sec.~\ref{sec:summary}, we summarize our presentation and lay out future
directions. Throughout this work, we assume a $\Lambda$CDM cosmology with
$\Omega_m = 0.27$, $\Omega_\Lambda = 0.73$, $\Omega_b = 0.045$, $h = 0.7$,
$\sigma_8 = 0.8$, and $Y_\mathrm{He} = 0.24$. These values are consistent with
the \textit{WMAP}-9 year results \citep{hinshaw_etal2013}.

\section{Modeling Quasars as Radiation Sources}
\label{sec:pop}

\subsection{Radiation-hydrodynamic simulations}
When modeling helium reionization, we employ the RadHydro code, which includes
$N$-body, hydrodynamics, and radiative transfer calculations simultaneously. The
code includes a particle mesh (PM) solver for gravity calculations, a fixed-grid
Eulerian code for solving hydrodynamics, and radiative transfer solved by
performing ray-tracing. For more details on the hydrodynamics portion of the
simulation code, see \citet{trac_pen2004}. For more details regarding the
RadHydro code and its application to hydrogen reionization, see
\citet{trac_cen2007} or \citet{trac_etal2008}.

The simulation strategy we employ for our exploration of helium reionization
consists of two steps. First, a high-resolution $N$-body simulation is run for a
given set of initial conditions. Halos are found on-the-fly using the
friend-of-friends algorithm, and a corresponding catalog of spherical
overdensity halos are saved at even steps in cosmological time
\citep{trac_etal2015}. Then, using the same initial conditions, a
medium-resolution simulation using the RadHydro code is run. In order to provide
accurate sources of ionizing photons for the radiative transfer calculations, it
is necessary to convert the halo catalogs produced from the first simulation
into quasar catalogs for the second simulation. Since the resolution of the
RadHydro simulations is comparatively low (typically a hydro grid unit is 10-100
$h^{-1}$kpc), the simulations are not able to accurately capture the subgrid,
galaxy-level physics to include quasars directly. Thus, either a halo-level
scaling relation or observational constraint is needed in order to create a
physically reliable sample. Rather than having to rely on scaling relations that
require several steps to convert between halo mass and quasar luminosity, we use
abundance matching to calculate luminosity as a function of mass, and then use
observations to create a population with the proper characteristics.

In order to calibrate the proper quasar properties to use, a suite of 10
$N$-body P$^3$M simulations with $L~=~1$~$h^{-1}$~Gpc and $2048^3$ dark matter
particles were run, which corresponds to a particle mass of
$m_p = 8.72 \times 10^9\ h^{-1}M_\odot$. The total volume is thus 10
($h^{-1}$~Gpc)$^3$; the BOSS measurement of the two-point correlation function in
\citet{white_etal2012} has an effective volume of 9.8 ($h^{-1}$~Gpc)$^3$, so the
volumes are comparable. Then halo finding was performed which produced the
associated halo catalog snapshot every 20 Myr between $2 \leq z \leq 10$. Since
only comparatively massive halos serve as hosts for the bright quasars of
interest, the simulations have a sufficient resolution to capture the required
number of halos.

\subsection{Quasar light curves}
\label{sec:light_curve}
The first step in our model construction is to define the properties of
individual quasars. The two most important of these are the light curve
(\textit{i.e.}, $L(t)$) and the quasar lifetime. The most common model found in
the literature for the light curve of quasars is the so-called lightbulb model,
in which a quasar emits radiation at a constant luminosity for a lifetime $t_q$
before turning off. Though largely unphysical, this model has the convenience of
being simple to implement in calculations. A further simplification is typically
made in which it is assumed that $t_q$ is independent of luminosity, so that
this quantity becomes a universal property.

A more realistic model of the light curve is to assume an exponential form. This
type of model can be motivated physically by noting that it corresponds to
Eddington accretion onto the central SMBH. Several variations on this version
include an exponential ramp-up to some peak luminosity followed by abrupt
turn-off, a symmetric exponential about some peak luminosity, or an exponential
ramp-up with a power-law fall-off in luminosity
\citep{hopkins_hernquist2009,mcquinn_etal2009}. While these models are more
physically motivated, they are slightly more complicated. The approach we
outline below is able to reproduce a given luminosity function for quasar light
curves of this form.

Specifically, we consider here two classes of quasar light curves: the
``lightbulb'' model and ``exponential'' model, defined as:
\begin{gather}
L_\mathrm{lb} (t) = L_\mathrm{peak} \Theta(t + t_q/2 - t_0)\Theta(t_q/2 - t + t_0), \label{eqn:lb} \\
L_\mathrm{exp} (t) = L_\mathrm{peak} \exp(-\abs{t_0 - t}/\tau), \label{eqn:exp}
\end{gather}
where $\Theta(t)$ is the Heaviside theta function and $t_0$ is the time when the
quasar reaches its peak luminosity $L_\mathrm{peak}$. In the exponential case,
the parameter $\tau$ can be treated as a free parameter in a manner analogous to
$t_q$ in the lightbulb case. Nevertheless, we relate $\tau$ to $t_q$, which we
will describe in more detail in Sec.~\ref{sec:am}.

Another consideration is the quasar lifetime itself, which in general need not
be a universal property of all quasars. We have parameterized quasar lifetime as
a function of luminosity using a power-law form:
\begin{equation}
t_q(L) = t_0 \qty(\frac{L}{10^{10}L_\odot})^\gamma,
\label{eqn:tq}
\end{equation}
where we vary the values of $t_0$ and $\gamma$. We explore models in which
$10^7 \leq t_0 \leq 10^9$ yr, and $-0.25 \leq \gamma \leq 0.10$.  Positive
values of $\gamma$ imply that brighter quasars have longer lifetimes compared to
dimmer ones, and $\gamma=0$ is the case of a universal lifetime for all quasars.

\subsection{Triggering rate}
\label{sec:tr}
We have discussed considerations for the individual quasars (\textit{i.e.},
light curves and lifetimes), and we wish to connect them to the universal quasar
population (\textit{i.e.}, the QLF). In order to do so, we use the concept of a
triggering rate $\dot{n}(L_\mathrm{peak},z)$, which dictates the differential
number density of quasars that reach their peak luminosity $L_\mathrm{peak}$ as
a function of luminosity and redshift per unit logarithmic luminosity. Using the
formalism outlined in \citet{hopkins_etal2006}, we distinguish between the peak
luminosity of a quasar $L_\mathrm{peak}$ and the instantaneous luminosity at
which it is measured for the construction of the QLF $L$, and relate the two
with the triggering rate $\dot{n}$. Essentially, the triggering rate must be
convolved with the light curve of the quasars, since the measured luminosity
function reflects a given quasar's current luminosity $L$ rather than its
intrinsic peak luminosity $L_\mathrm{peak}$. The result of this convolution is
the observed QLF from the intrinsic triggering rate:
\begin{equation}
\phi(L,z) = \int \dv{t(L,L_{\mathrm{peak}})}{\log L} \dot{n}(L_{\mathrm{peak}},z) \dd{\log L_{\mathrm{peak}}}.
\label{eqn:ndot}
\end{equation}
As explained in \citet{hopkins_etal2006}, $\phi(L)$ is the QLF (\textit{i.e.},
the comoving number density of quasars per logarithmic bin in luminosity), and
the quantity $\dv*{t(L,L_{\mathrm{peak}})}{\log L}$ is the amount of time that a
quasar spends in a logarithmic luminosity bin. Essentially, the triggering rate
can be thought of as analogous to the halo mass function, though with the light
curve convolution to account for changes in quasar brightness. In simple cases
of the light curve the triggering rate can be solved for analytically: in the
case of a lightbulb light curve, $\dv*{t(L,L_{\mathrm{peak}})}{\log L}$ is a delta
function at $L=L_\mathrm{peak}$, and so the triggering rate is proportional to
the QLF:
\begin{equation}
\dot{n}_{\mathrm{lightbulb}}(L,z) = \frac{1}{t_q} \phi(L,z).
\label{eqn:ndotlb}
\end{equation}
In the case of an exponential light curve as defined in
Eqn.~(\ref{eqn:exp}), we have
\begin{equation}
\dot{n}_{\mathrm{exp}}(L,z) = \frac{1}{2\tau} \eval{\dv{\,\phi(L,z)}{\log L}}_{L=L_\mathrm{peak}},
\label{eqn:ndotexp}
\end{equation}
where the factor of 2 arises because a quasar will be observed at a luminosity
$L$ while its luminosity is increasing and then decreasing. In practice, the QLF
is typically reported in magnitude units rather than luminosity. One common
convention is to report the quasar's absolute $i$-band magnitude at $z=2$,
$M_i(z=2)$. This quantity is then converted to the specific luminosity at 2500
\AA, $L_{2500\ \mathrm{\AA}}$, in cgs units (erg s$^{-1}$ Hz$^{-1}$) by using
Eqn.~(4) of \citet{richards_etal2006}:
\begin{multline}
\log_{10}\qty(\frac{L_{2500\ \mathrm{\AA}}}{4\pi d^2}) = \\
          -0.4 \qty[M_i(z=2) + 48.60 + 2.5 \log_{10}(1+2)],
\label{eqn:L2500}
\end{multline}
where $d = 10\ \mathrm{pc} = 3.08\times 10^{19}$ cm. To find the approximate
bolometric luminosity, the relation of \citet{shen_etal2009} can be used to
convert $M_i(z=2)$ to luminosity in erg s$^{-1}$:
\[
M_i(z=2) = 90 - 2.5 \log_{10}(L).
\]
One should note that this relation is approximate, and depends on the assumed
spectral energy distribution (SED) of the quasar. Eqn.~(\ref{eqn:ndot}) is
soluble for a few classes of light curves, such as the ones explored here.

\subsection{Abundance matching}
\label{sec:am}
The technique of abundance matching has already been applied to populations of
galaxies with great success (\textit{e.g.},
\citealt{simha_etal2012,hearin_etal2013}), and has also been discussed in the
context of quasars (\textit{e.g.},
\citealt{martini_weinberg2001,porciani_etal2004,croton_2009}). However, we wish
to extend the techniques mentioned above to include different quasar light
curves and lifetimes. The methods we outline below are also fairly general, and
can be extended to include semi-analytic models as well. We start with the
\textit{Ansatz} for abundance matching of galaxies, namely that the most
luminous galaxies are found in the most massive halos. This makes intuitive
sense: more massive halos have more dark matter and baryonic matter to
eventually convert to stars. Specifically, halo mass is highly correlated to the
luminosity in the red bands, which shows the percentage of older stellar mass.

For quasars, we have a similar situation where the most luminous quasars are
found in the most massive halos. However, in this case the situation is slightly
more complicated because quasars have a lifetime which is much shorter than the
period from the halo's formation to the activation of the quasar. Thus, we need
to introduce a factor to account for the fact that not all halos host an active
quasar. If we assume that the fraction of halos hosting an active quasar is
universal (\textit{i.e.}, independent of halo mass or quasar luminosity), we can
express abundance matching for quasars, assuming a lightbulb light curve, as:
\begin{equation}
\phi(>L) = f_\mathrm{on} n_\mathrm{halo} (>M).
\label{eqn:amcum}
\end{equation}
Expressed this way, $f_\mathrm{on}$ is simply the fraction of halos of a mass
$M$ that host an active quasar. Alternatively, we could define this fraction in
terms of the quasar lifetime:
\begin{equation}
f_\mathrm{on}(L,z) = \frac{t_q(L)}{t_H(z)},
\end{equation}
where in some models $t_H(z)$ is formulated as the halo lifetime
\citep{martini_weinberg2001}, or the Hubble time \citep{conroy_white2013}. We
follow \citet{conroy_white2013} and use the Hubble time. As we shall see,
though, the exact choice for $t_H(z)$ does not strongly affect the results. For
the redshifts of interest, for a uniform value of $t_q = 30$ Myr, this implies
that $f_\mathrm{on} \sim 0.1-1\%$.

We can generalize the procedure of abundance matching to different light curves
by using the triggering rate. In integral form, we can write abundance matching
as equating the cumulative number of quasars above a particular peak luminosity
given by the triggering rate with the cumulative number of halos given by the
halo mass function. The total number of halos which should host quasars within a
time interval $\Delta t$ is:
\begin{multline}
\int_{\Delta t} \int_L^\infty \dot{n}(L^*) \dd{\log L^*} \dd{t} \\
\begin{aligned}
&= \int_{\Delta t} \int_L^\infty \dv{n_\mathrm{halo} (L^*)}{\log M^*} \dv{\log M^*}{\log L^*} \dv{P}{t} \dd{\log L^*} \dd{t}  \\
&= \frac{\Delta t}{t_H} \int_M^\infty \dv{n_\mathrm{halo}(M^*)}{\log M^*} \dd{\log M^*}. \label{eqn:am}
\end{aligned}
\end{multline}
This form of our abundance matching equation becomes the central mechanism by
which we are able to equate quasar luminosity with host halo mass. In this
construction, we have implicitly used the mass-to-light ratio
$\dv*{\log M}{\log L}$ to convert halo mass to quasar luminosity. Additionally,
we have introduced the factor $\dv*{P}{t}$ to represent the probability that an
individual halo will host a quasar. We have set this quantity to be equal to
$1/t_H$. Thus, for the case of a lightbulb light curve and a universal quasar
lifetime, this formalism reduces to Eqn.~(\ref{eqn:amcum}). Formally, this
expression is an expansion of $\dot{n}(L^*, z)$ about $z$ that is first-order
accurate to $\Delta t/t_H$ \citep{hopkins_etal2006}. Thus, so long as the
time-steps between determining the triggering rate are small compared to $t_H$
(defined either as the Hubble time or the halo lifetime, both several orders of
magnitude longer than the typical quasar lifetime), this expression should
reproduce the target QLF.

In the exponential case, we are free to choose the parameter $\tau$ in any way
that we like, as long as it is constant with respect to $L$ (though it may vary
with $L_\mathrm{peak}$). We have chosen $\tau$ such that
$\dot{n}(L_\mathrm{peak})$ is the same between the lightbulb and exponential
cases for all luminosities. We accomplish this by equating
Eqn.~(\ref{eqn:ndotlb}) and Eqn.~(\ref{eqn:ndotexp}), and solving for $\tau$ in
terms of $t_q$. The expression involves the ratio of the QLF and its
derivative. This means that when we perform abundance matching, the same
implicit mass-to-light ratio is used in the two cases. Since the halo mass
function is the same between the two cases (due to the same population of halos
being used), and the functional form of $\dot{n}$ is the same, we must have the
same form of $\dv*{\log M}{\log L}$. This has the advantage of allowing us to
apply certain intuition from the lightbulb case to the less straightforward
exponential case. The downside to this approach is that when exploring the
parameter space of quasar lifetimes $t_q$ in the lightbulb case, it is not
immediately obvious how this translates to the exponential time constant $\tau$,
since we effectively have different values of $t_q$ for different
luminosities. For instance, even in cases where $t_q$ is independent of
luminosity, $\tau$ still changes as a function of $L$. However, the benefits of
being able to interpret the results of the exponential case using the intuition
provided by the lightbulb case outweigh the downsides of not exploring
parameters in $\tau$ directly.

The general procedure is as follows. 
\begin{enumerate}
\item The halo mass found from the halo catalog at redshift $z_\mathrm{cat}$ is
  read and converted to an expected number density in a particular cosmology
  using the universal mass function described in \citet{tinker_etal2008}. The
  fitted form of the mass function is used rather than the empirical one from
  the catalog in order to decrease the variation in number density at the
  high-mass end, since these quasars are disproportionately important for the
  reionization process.
\item Using Eqn.~(\ref{eqn:am}), the halo number density is converted to an
  expected quasar number density using a specified QLF.
\item The quasar magnitudes are binned into equal intervals in magnitude
  $\Delta M$, such that the expected triggering rate
  $\dot{n}(M,z_\mathrm{cat})\ldots\dot{n}(M+\Delta M,z_\mathrm{cat})$ is found,
  which is converted from a number density to a total number $\dot{N}(M)$ using
  the volume of the simulations.
\item Within each magnitude bin, each quasar is assumed to have an equal
  probability of becoming active. Each quasar candidate is randomly turned on
  with probability $1/\dot{N}_\mathrm{bin}(M)$.
\item To ensure that the volume self-consistently follows the merging of the
  underlying host halos, the quasars are propagated forward using a halo merger
  tree. By design, the halo catalog snapshots are made at times that are shorter
  than the expected lifetimes of the quasars. This approach allows for halos
  hosting quasars to be tracked throughout the simulation. In most cases, an
  active quasar from time step $i-1$ in a progenitor halo passes to the single
  descendent halo at time step $i$. Additionally, this halo hosting an active
  quasar is not eligible to host a new quasar. This approach covers the majority
  of halos for the majority of time steps. However, there are several special
  cases related to merger events worth discussing. Specifically, when two
  progenitor halos merge into a single descendent and one of them is hosting an
  active quasar, the descendent halo inherits the active quasar. If a single
  active progenitor halo splits to form two descendent halos, the larger halo
  retains the quasar. In the case of a merger between two active quasar halos,
  only the larger quasar survives. These cases represent a comparatively few
  number of instances of our total population evolution, and do not strongly
  influence our conclusions.
\end{enumerate}

\begin{deluxetable*}{ccllllll}
\tablecaption{A list of the QLF parameters of the Datasets Incorporated. \label{table:qlf}}
\tablewidth{0pt}
\tablehead{\colhead{Dataset} & \colhead{$z$} & \colhead{$\log_{10}(\phi^*)$\tablenotemark{a}} & \colhead{$M_0^*$\tablenotemark{b}} & \colhead{$k_1$\tablenotemark{c}} & \colhead{$k_2$} & \colhead{$\alpha$} & \colhead{$\beta$}}
\startdata
R13 & 2.2-3.5 & $-5.93^{+0.02}_{-0.01}$ & $-26.57^{+0.04}_{-0.02}$ & $-0.689^{+0.021}_{-0.027}$ & $-0.809^{+0.033}_{-0.166}$ & $-1.29^{+0.15}_{-0.03}$ & $-3.51^{+0.09}_{-0.18}$ \\
M12 & 3.2 & $-6.58^{+0.26}_{-0.79}$ & $-27.03 \pm 0.68$ & \multicolumn{1}{c}{\ldots} & \multicolumn{1}{c}{\ldots} & $-1.73 \pm 0.11$ & $-2.98 \pm 0.21$ \\
M12\tablenotemark{d} & 4 & $-7.12^{+0.62}$ & $-27.13 \pm 2.99$ & \multicolumn{1}{c}{\ldots} & \multicolumn{1}{c}{\ldots} & $-1.72 \pm 0.28$ & $-2.6 \pm 0.63$ \\
M13\tablenotemark{e} & 5 & $-8.47^{+0.20}_{-0.24}$ & $-28.70^{+0.27}_{-0.33}$ & \multicolumn{1}{c}{$-0.47$} & \multicolumn{1}{c}{\ldots} & $-2.03^{+0.15}_{-0.14}$ & $-4.00$\\
M13 & 5 & $-7.63^{+0.30}_{-0.25}$ & $-27.34^{+0.60}_{-0.49}$ & \multicolumn{1}{c}{$-0.47$} & \multicolumn{1}{c}{\ldots} & $-1.50$ & $-3.12^{+0.28}_{-0.41}$\\
M13 & 5 & $-7.93^{+0.03}_{-0.03}$ & $-27.88$ & \multicolumn{1}{c}{$-0.47$} & \multicolumn{1}{c}{\ldots} & $-1.80$ & $-3.26$\\
\enddata
\tablenotetext{a}{$\phi^*$ has units of Mpc$^{-3}$ mag$^{-1}$.}
\tablenotetext{b}{$M_0^* = M_i(z=2) = M_{1450}-1.486$.}
\tablenotetext{c}{$k_1$
  and $k_2$ are defined for models with redshift evolution in
  Eqns.~(\ref{eqn:ledephi}-\ref{eqn:ledephi2}).}
\tablenotetext{d}{The authors
  of M12 provide a value for $\phi_0^*$ at $z \sim 4$ where the reported error
  is greater than the value itself. Since this value must be positive, the
  resulting lower-bound is unphysical. We reproduce the value and upper-bound
  here for completeness, but do not include this value directly when determining
  the values of the QLF. See Appendix~\ref{sec:appendixa} for further details.}
\tablenotetext{e}{In M13, the authors provide three fits, each with at least one
  parameter held constant. Values without error ranges indicated correspond to
  the parameters held fixed for a particular fit.}
\end{deluxetable*}

\subsection{The quasar luminosity function}
\label{sec:qlf}
Throughout this work, we use a series of QLFs as determined at different
epochs. For relatively low-redshift ($2 \lesssim z \lesssim 3$), we use the QLFs
as determined by R13 from the BOSS survey, specifically the high-$z$ stripe 82
sample (S82) form which includes luminosity evolution and density evolution
(LEDE). Above a redshift of 3, the QLF has been measured at $z \sim 3.2$ and
$z \sim 4$ by M12 using data from COSMOS. \footnote{Additionally, the QLF at
  $z \sim 4$ has also been measured by \citet{glikman_etal2011} and
  \citet{ikeda_etal2011}. As noted in M12, the normalization of the QLF of
  \citet{ikeda_etal2011} is comparable, whereas the normalization of
  \citet{glikman_etal2011} is larger than the others by a factor of $\sim 4$.
  M12 notes that the difference can be caused by contamination of the
  faintest-magnitude bins from dwarf stars and high-redshift galaxies. In the
  following analysis, we use the results from M12.} At $z \sim 5$, the QLF has
been measured by M13 using data from the SDSS. \footnote{An upper limit for the
  QLF at $z\sim 5$ was found by \citet{ikeda_etal2012}, which is consistent with
  the results of M13.} Although these works use slightly different values for
cosmological parameters from the ones assumed here, the impact on the reported
quantities is minimal.

In order to span the different epochs over which the luminosity function has
been measured, it is necessary to combine the different data sets. All of the
data sets fit to a double power law form of the QLF, written as:
\begin{equation}
\Phi(M) = \frac{\phi^*}{10^{0.4(1+\alpha)(M-M^*)} + 10^{0.4(1+\beta)(M-M^*)}},
\label{eqn:qlf}
\end{equation}
where $\Phi$ is the comoving number density of quasars of magnitude $M$ per unit
magnitude, $\phi^*$ is the normalization of the QLF, $\alpha$ is the faint-end
slope of the luminosity function, $\beta$ is the steep-end slope (which is
reversed from the parameterizations of M12), and $M^*$ is the so-called break
magnitude where the luminosity function transitions from the faint-end to the
steep-end. In most formulations at high-redshift, redshift evolution is
incorporated by a change in $\phi^*$, $M^*$, or both, that is linear in
redshift. For the data from R13, the evolution is given by the equations:
\begin{align}
\log_{10} \phi^*(z) &= \log_{10} \phi^*_0 + k_1(z-2.2), \label{eqn:ledephi} \\
M_i^*(z) &= M^*_0 + k_2(z-2.2). \label{eqn:ledem}
\end{align}
For the data in M13, there is linear evolution in $\log_{10} \phi^*$ as well, given as:
\begin{equation}
\log_{10} \phi^*(z) = \log_{10} \phi^*_0 + k_1(z-6).
\label{eqn:ledephi2}
\end{equation}

To combine the R13, M12, and M13 data sets into a single set of quantities, we
first assume that the results from R13 are accurate for redshifts $z \leq 3.5$.
This is the nominal limit of the LEDE fits, and though there are small
differences between the fit QLF and the binned data, overall the fits are
excellent. To incorporate the results at higher redshifts, we cast the four
parameters of the QLF ($\phi^*$, $M^*$, $\alpha$, and $\beta$) as quantities
that have linear evolution in redshift. We define these parameters as:
\begin{subequations}
  \begin{align}
    \log_{10} \phi^*(z) &= \log_{10}\phi^*_0 + c_1 (z-3.5), \label{eqn:logphiz} \\
    M^*(z) &= M^*_0 + c_2(z-3.5), \label{eqn:miz} \\
    \alpha(z) &= \alpha_0 + c_3(z-3.5), \label{eqn:alphaz}\\
    \beta(z) &= \beta_0 + c_4(z-3.5). \label{eqn:betaz}
  \end{align}
\end{subequations}
These parameterizations are applied to redshifts where $z > 3.5$. The constant
values are defined to be equal to the values of R13 at $z=3.5$, and the values
for the slopes ($c_1$--$c_4$) are allowed to take on a range of values. The range
is generally chosen such that the values for the different parameters brackets
the range of best-fit values provided by the highest redshift (M13) data. The
fiducial values for the slopes are taken to be ones that reasonably reproduce
the high-redshift measurements. Table~\ref{table:qlf} shows the fiducial values
for the slopes, as well as the range of values for the parameters at $z \sim 5$
used in the parameter space exploration in Sec.~\ref{sec:reion}. For a complete
discussion on selecting the parameters for the QLF, see
Appendix~\ref{sec:appendixa}.

Table~\ref{table:qlf} lists the parameters that we include from the measurements
of R13, M12, and M13. The values from M12 are not included in the fitting
procedure directly, and serve primarily as a consistency check due to their
comparatively large error bars. The parameters from M13 are determined at
$z \sim 5$, and the ones from M12 are determined at $z \sim 4$ and $z \sim 3.2$.
Note that the authors of M13 provide three independent fits to their data, which
are all incorporated into the final QLF parameterization. (See
Appendix~\ref{sec:appendixa} for more details.) For the measurements from R13,
whose fiducial LEDE model includes redshift evolution in $\phi^*$ and $M^*$, the
model is valid over a range of redshift, from $2.2 \leq z \leq 3.5$. For the
purposes of generating our quasar catalogs, we are interested in exploring the
QLF until $z=2$. For the sake of simplicity, we simply extend the LEDE model
from R13 to this redshift. Although the LEDE fit is ostensibly not valid below
$z=2.2$, we expect helium reionization to be largely finished by this redshift,
and so the precise form of the QLF at $z \sim 2$ is not of fundamental
importance to our study. Also, for the value of $M^*$, it is necessary to
convert to a single magnitude system. As explained in Sec.~\ref{sec:tr}, we use
$M_i(z=2)$, the absolute $i$-band magnitude at $z=2$. The QLFs of M12 and M13
use $M_{1450}$, which is related to $M_i(z=2)$ by $M_i(z=2) = M_{1450} - 1.486$
(\citealt{richards_etal2006}; \citealt[][Appendix B]{ross_etal2013}). Note that
this conversion assumes that the quasar SED follows a power-law with an
effective spectral index of $\alpha = 0.5$ (using the convention that
$f_\nu(\nu) \propto \nu^{-\alpha}$). Modifying the spectral index $\alpha$
changes the magnitude conversion, so care must be taken when converting between
magnitude systems. See Appendix~\ref{sec:appendixa} for further discussion.

\begin{deluxetable}{ccc}
  \tablecaption{A List of the Parameters Used in Eqns.~(\ref{eqn:logphiz}-\ref{eqn:betaz}) Based on the Data Listed in Table~\ref{table:qlf}. \label{table:parameters}}
  \tablewidth{0pt}
  \tablehead{\colhead{Parameter} & \colhead{Fiducial Value} & \colhead{Parameter Range}}
  \startdata
  $\log_{10}\phi^*_0$ & $-6.82$ & $\cdots$ \\
  $c_1$ & $-0.790$ & $[-1.10, -0.536]$ \\
  $M^*_0$ & $-27.6$ & $\cdots$ \\
  $c_2$ & $-0.238$ & $[-0.716, 0.170]$ \\
  $\alpha_0$ & $-1.29$ & $\cdots$ \\
  $c_3$ & $-0.324$ & $[-0.493, -0.140]$ \\
  $\beta_0$ & $-3.51$ & $\cdots$ \\
  $c_4$ & $0.0333$ & $[-0.327, 0.260]$
  \enddata
  \tablecomments{These provide a fit to the luminosity function through
    redshift, and ensure that the abundance of quasars matches observations as
    nearly as possible. For additional details on the parameters and the fitting
    procedure, see Appendix~\ref{sec:appendixa}.}
\end{deluxetable}

\begin{figure}[t]
  \centering
  \includegraphics[width=0.45\textwidth]{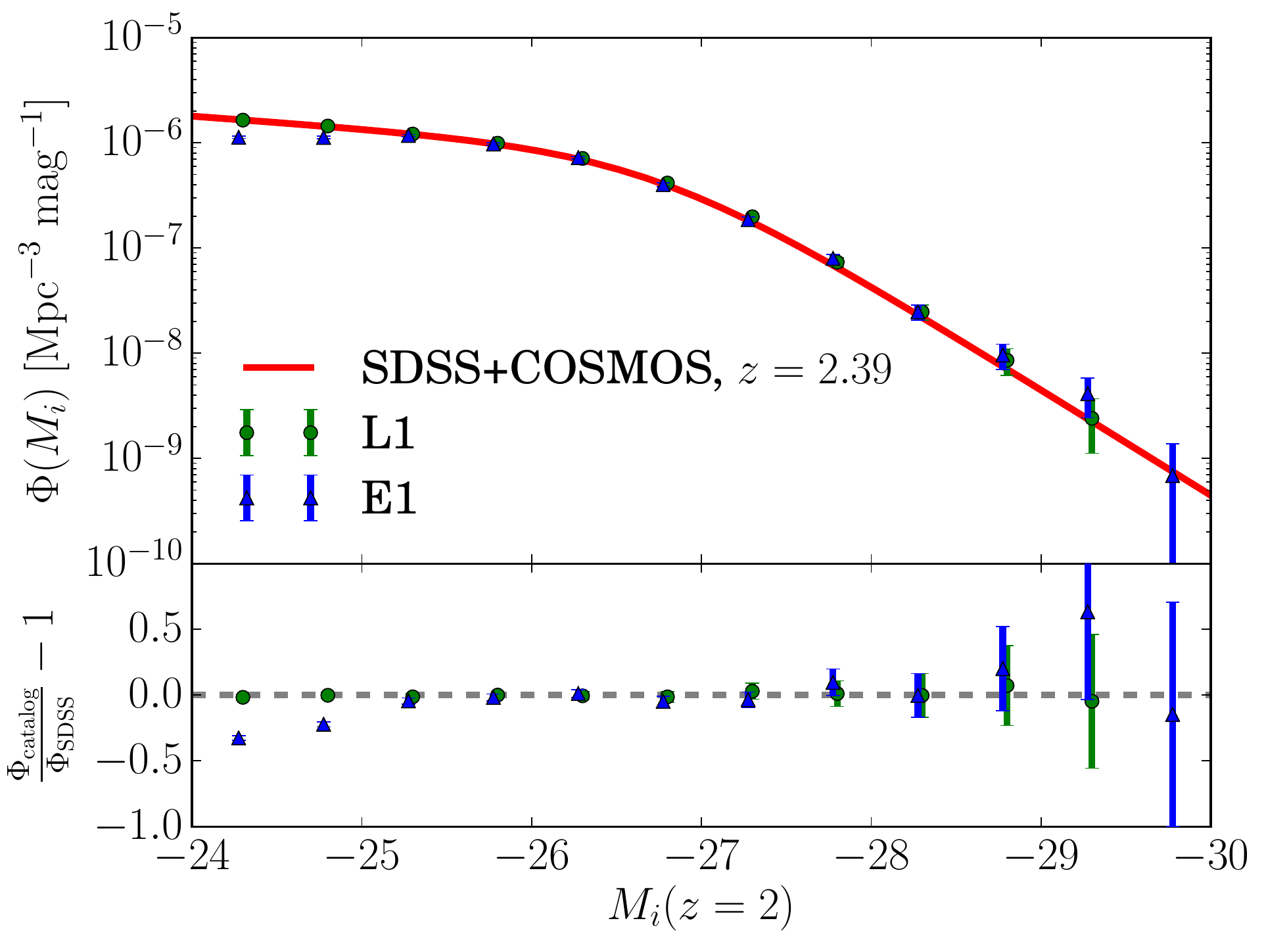}
  \caption{A comparison of the composite quasar luminosity functions from the
    SDSS+COSMOS measurements
    \citep{masters_etal2012,mcgreer_etal2013,ross_etal2013} to our abundance
    matching method, plotted with Poisson error bars. The two different quasar
    models (defined in Table~\ref{table:models}) are offset from each other for
    visual clarity. The agreement is excellent for comparatively dim quasars
    which are more common, but there is some discrepancy for bright objects. The
    reason for this disagreement is primarily due to Poisson noise, since these
    objects are rare even for the large (1 ($h^{-1}$~Gpc)$^3$) simulation
    volume. At low luminosity in the exponential case, the completion limits of
    dark matter halo hosts at this mass become noticeable. See Sec.~\ref{sec:am}
    for further discussion.}
  \label{fig:qlf}
\end{figure}

Figure~\ref{fig:qlf} shows the combined QLF from R13, M12, and M13 (which at
this epoch is essentially that of R13), as well as two different quasar models
at $z\sim 2.4$. We can see that there is generally very good agreement between
the constructed quasar catalog and the target luminosity function, as should be
expected. The differences between the constructed catalogs and target luminosity
function are typically on average $\lesssim 5\%$, which is comparable to or
smaller than the uncertainties in the luminosity function itself at these
redshifts. At high luminosities ($M_i \lesssim -28$), though, there are some
comparatively large differences that can arise between the predicted and
empirical luminosity functions. This deviation is largely due to Poisson
shot-noise introduced by the rarity of the objects. For objects in this
luminosity range, there are typically only a few objects ($\order{10}$) in the
entire 1 ($h^{-1}$~Gpc)$^3$ volume. At the dim end of the QLF, there can be
insufficient halos of a particular mass given the mass resolution of our
simulation. The minimum halo mass is
$M_\mathrm{halo,min} = 4.36 \times 10^{11}\ h^{-1}M_\odot$. Since quasars with
$M_i \leq -25$ are most important for this study, this does not affect our
results significantly.

Throughout most of the following analysis, we focus our attention on several
models in particular, parameterized in terms of $t_0$ and $\gamma$ as in
Eqn.~(\ref{eqn:tq}). The first four of these models have particularly good
agreement with the BOSS measurements. The last two are included to demonstrate
how the clustering signal changes as a function of $t_0$ for a fixed value of
$\gamma$: models L1, L3, and L4 all have the same $\gamma$ value. We summarize
these models in Table~\ref{table:models}.

\section{Clustering Measurements}
\label{sec:clustering}

\subsection{Two-point Correlation Function}
\label{sec:2pcf}
By construction, our method matches the input QLF at all redshifts, regardless
of the individual properties of the underlying quasar population. However, we
are not guaranteed to match the observed clustering of quasars. Changing the
implicit mass-to-light ratio of Eqn.~(\ref{eqn:am}) through changing the quasar
lifetimes will affect how halos are populated with quasars. In general, longer
quasar lifetimes lead to quasars of the same luminosity being matched into hosts
of larger masses. Since their hosts are more biased, this leads to quasars of
the same luminosity showing a larger clustering signal. This is true at all
luminosities. We want to match the clustering because it can affect the topology
of reionization. There can also be spatial correlations present in the radiation
field as a result of reionization, which are important for making measurements
of the BAO from the Lyman-$\alpha$ forest (\textit{e.g.},
\citealt{white_etal2010,slosar_etal2013}).

Here, we explore how to include clustering measurements from the two-point
correlation function in our quasar catalog. Recent results from the BOSS survey
for the clustering of quasars in the redshift range of interest are presented in
\citet{white_etal2012}. The above work examines the clustering signal of quasars
in both 2D-projected and 3D-redshift-space correlation functions at intermediate
scales ($3 \lesssim s \lesssim 25$ $h^{-1}$Mpc). The authors also introduce
luminosity cuts to make the results more robust. For the purposes of this
comparison, we consider their selection for which they imposed luminosity cuts
on both the bright and faint ends, so that only objects with
$-25 \geq M_i \geq -27$ were considered across the entire redshift range
(Sample 4 as defined by the authors). For a fair comparison, we impose similar
cuts on our object selection. We also examine the redshift evolution of the
results, and compare against the high-$z$/low-$z$ samples (Samples 5 and 6) as
well. See Appendix~\ref{sec:appendixb} for further discussion of these different
redshift samples.

\begin{deluxetable}{cccc}
\tablecaption{A List of the Parameters of Some Quasar Models Considered. \label{table:models}}
\tablewidth{0pt}
\tablehead{\colhead{Model Name} & \colhead{Light Curve} & \colhead{$\log_{10}(t_0/\mathrm{yr})$\tablenotemark{a}} & \colhead{$\gamma$}}
\startdata
L1 & Lightbulb & 7.75 & 0 \\
L2 & Lightbulb & 8.25 & $-0.125$ \\
E1 & Exponential & 7.25 & 0 \\
E2 & Exponential & 7.75 & $-0.15$ \\
L3 & Lightbulb & 7 & 0 \\
L4 & Lightbulb & 8.5 & 0 \\
\enddata
\tablenotetext{a}{$t_0$ and $\gamma$ as defined in Eqn.~(\ref{eqn:tq}).}
\end{deluxetable}

We explore the parameter space of available quasar models by examining the
lightbulb and exponential light curves defined in Eqns.~(\ref{eqn:lb}) and
(\ref{eqn:exp}), as well as luminosity-dependent quasar lifetimes defined in
Eqn.~(\ref{eqn:tq}), parameterized by $t_0$ and $\gamma$. For each combination
of parameters, we construct a quasar catalog in the manner described
above.\footnote{There are several extreme models where the number of objects is
  significantly fewer than the number predicted by the quasar luminosity
  function. This is not a failure of our methodology, but rather instances of
  there being too few halo objects of a given mass to host quasar objects. In
  essence, $f_\mathrm{on}$ is so small that we reach the resolution limits of
  the simulation. In these cases, we add particles from a second-order
  Lagrangian perturbation theory (2LPT) simulation of the same initial
  conditions at the same redshift in order to define a set of ``random''
  particles that are still representative of the underlying matter
  distribution. We randomly sample from these particles in order to fill out the
  catalog to the expected number. This ensures that we do not measure a
  statistically significant clustering measurement when the catalog is clearly
  unphysical.} Then, we extract from this catalog all objects that satisfy the
magnitude constraints at the central redshift of the survey $z = 2.39$. This
redshift represents the average redshift of quasars chosen in the BOSS sample;
the actual quasar objects span in redshift from $2.2 < z < 2.8$. However, as
noted in \citet{white_etal2012}, the redshift evolution of the signal is
weak. Thus, extracting objects from our quasar catalogs at a single redshift
rather than a range should have little effect on our overall conclusions. We
measure the monopole of the two-point correlation function using the ``natural
estimator'' $\xi$:
\begin{equation}
\xi(s) = \frac{\ev{DD(s)}}{\ev{RR(s)}} - 1,
\end{equation}
where $\ev{DD(s)}$ is the average number of quasar pairs from the quasar catalog
separated by a real-space distance of $[s - \Delta s/2,s + \Delta s/2]$, and
$\ev{RR(s)}$ is the number of pairs of points at the same separation drawn from
a distribution with Poisson noise.

\subsection{Calculating $\chi^2$ values}
\label{sec:chi2}

\begin{figure}[t]
  \centering
  \includegraphics[width=0.45\textwidth]{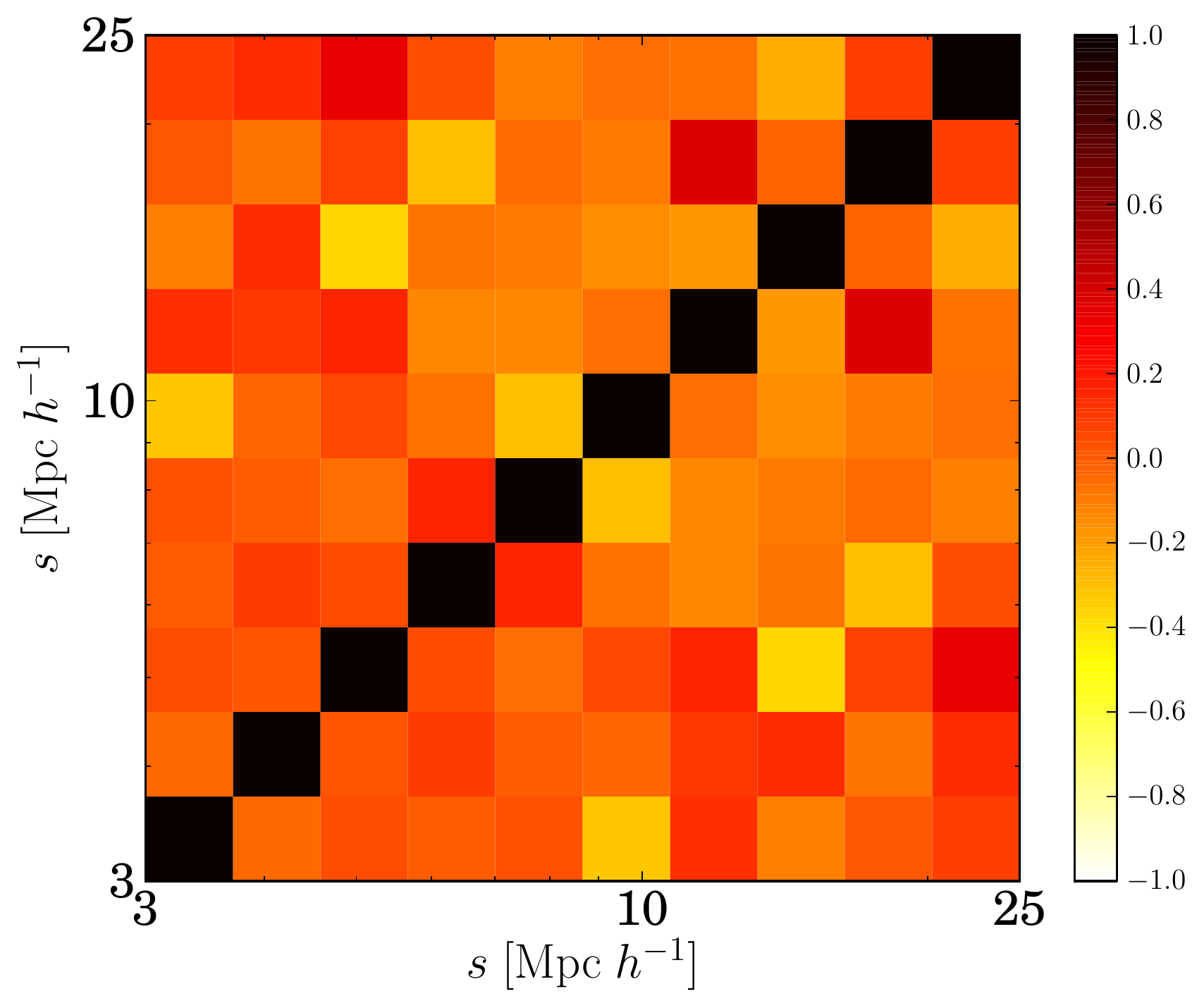}
  \caption{The correlation matrix for the L1 model. Note how the matrix is
    dominated by the diagonal entries, which is to be expected for shot-noise
    dominated measurements. The small off-diagonal terms suggest that the
    covariance matrix has converged numerically, and should be stable when
    inverting. This type of structure is seen in all models considered.}
  \label{fig:cor}
\end{figure}

In order to quantify the statistical uncertainty in our catalog, we ran a suite
of 10 $N$-body simulations with different initial conditions. We then performed
our abundance matching procedure on each of the different simulations, including
several realizations for each volume. Since our abundance matching procedure
stochastically determines which halos should be hosting active quasars at a
given time step, we create several quasar catalogs for each individual halo
catalog, using a different initial random seed (three realizations per volume
for these results). Additionally, we have augmented the effective number of
samples by including redshift space distortions along the different principal
axes of the simulation. This strategy gives us a total of 90 samples for which
to measure the clustering signal. The best estimate for the correlation function
$\xi(s)$ for a given radial bin $s_i$ is given by averaging over all of the
individual estimates $\xi_k$:
\begin{equation}
\bar{\xi}(s_i) = \frac{1}{N} \sum_{k=1}^N \xi_k(s_i)
\end{equation}
We then estimate the covariance between the radial bins by computing the entries
of the covariance matrix $C_{ij}$. We compute the entries of the covariance
matrix as \citep{zehavi_etal2005}:
\begin{equation}
C_{ij} = \frac{1}{N} \sum_{k=1}^N \qty(\xi_k(s_i) - \bar{\xi}(s_i)) \qty(\xi_k(s_j) - \bar{\xi}(s_j)).
\end{equation}
The correlation matrix entries for our model L1 is plotted in
Figure~\ref{fig:cor}. Notice that the diagonal entries dominate, which means
that the bins are mostly independent of each other and dominated by shot-noise
\citep{valageas_etal2011,white_etal2012}. Implicitly, the samples have been
treated as being independent, and this is almost surely not the case. Although
the 10 volumes as a whole can be treated as being statistically independent, the
different realizations based on the same halo catalog are likely
correlated. Further, the projections of peculiar velocities along different axes
for the same realization are also likely to produce correlated results. However,
producing a sufficient number of independent realizations to decrease the noise
in the covariance matrix is computationally infeasible. Further, the variance in
the clustering signal among quasar catalog realizations for a given
$(t_0,\gamma)$ pair is comparable to small displacements in the $t_0$--$\gamma$
parameter space, so it is necessary to include this source of uncertainty. Since
we are interested only in finding models that are consistent with the BOSS
measurements which have their own set of observational uncertainties, we feel
that this approach produces sufficiently accurate results.

\begin{figure}[t]
  \centering
  \includegraphics[width=0.45\textwidth]{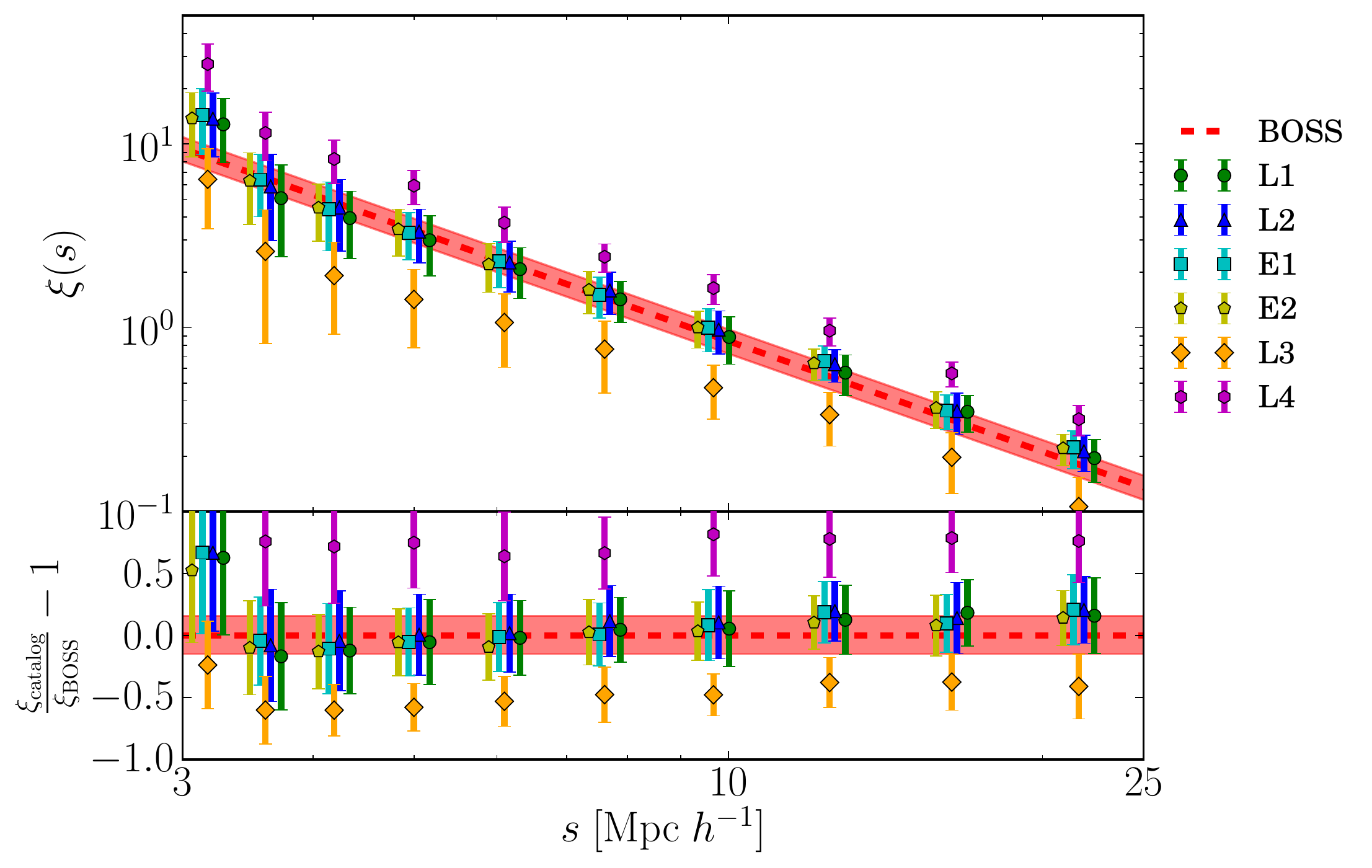}
  \caption{The quasar two-point correlation function from
    \citet{white_etal2012}, compared to several models whose parameters are
    described in Table~\ref{table:models}. All measurements were made at the
    same values of $s$, but are offset from each other for visual clarity. The
    shaded error regions on the measurements from BOSS are the reported
    1$\sigma$ error bars, and the error bars on the models are the square root
    of the diagonal elements of the covariance matrix. Note that for the same
    value of $\gamma$, increasing $t_0$ leads to a larger clustering signal
    (compare L3, L1, and L4 in order of increasing $t_0$). See the text for
    additional details.}
  \label{fig:clustering}
\end{figure}

Once the entries of the covariance matrix have been computed, the difference
vector $\delta(s_i) \equiv \xi_\mathrm{model}(s_i) - \xi_\mathrm{BOSS}(s_i)$ is
calculated. The correlation function $\xi_\mathrm{BOSS}$ is fit to a power law:
$\xi_\mathrm{BOSS}(s) = (s/s_0)^\beta$, where the authors have fixed the value
of $\beta=-2$. In order to investigate the impact this choice has on the
conclusions, we performed fits on the correlation function measured from our
quasar catalogs using two different parameterizations: one where the best-fit
value of $s_0$ was found when fixing $\beta=-2$, and another where the value of
$s_0$ and $\beta$ were both fit. In the length scales used for our analysis
($3 \leq s \leq 25$), the deviation of $\beta$ from the fiducial value of $-2$
was small, typically less than 5\%. Furthermore, the values for $s_0$ were also
largely similar between a fixed slope or a varying one, with deviations
typically less than 1\%. Thus, the choice to set $\beta=-2$ does not strongly
bias the results presented here, or the values reported in $\xi_\mathrm{BOSS}$.

When comparing one of the quasar models with the BOSS results, the $\chi^2$
value of the model is then given by:
\begin{equation}
\chi^2 = \delta^\mathrm{T} C^{-1} \delta.
\end{equation}
To define the model that fits the BOSS observations best, we want to minimize
the $\chi^2$ value of the model. A two-dimensional space in $t_0$ and $\gamma$
is constructed for both of the light curves, and this space is explored using
regular grid points. Following the analysis of \citet{white_etal2012}, a
$\chi^2$ distribution with nine degrees of freedom is assumed. Using this
distribution, the $\chi^2$ value for a particular model is converted to a
confidence interval. An equivalent $n\sigma$ value is computed based on the
confidence interval ($1\sigma$ if the enclosed probability is 0.683, $2\sigma$
if it is $0.955$, etc.). This statistic demonstrates how ``consistent'' a
particular model is with the BOSS observations.

Figure~\ref{fig:clustering} shows the clustering measurements for several of our
well-fitting models compared to the BOSS measurements. The values of these
models are given in Table~\ref{table:models}. In general, as $t_0$ increases at
a fixed value of $\gamma$, the clustering signal increases as well. Compare
specifically the L3, L1, and L4 models, which have the same value of $\gamma$
but have respectively increasing values of $t_0$. Mathematically, this behavior
can be seen from the form of Eqn.~(\ref{eqn:am}): for the same luminosity and
mass functions but a larger value of $f_\mathrm{on} \propto t_q$, quasars of the
same luminosity will shift to more massive host halos. Since the clustering
signal increases with the mass, it follows that increasing $t_0$ will increase
the clustering signal. For similar reasons, increasing values of $\gamma$ for
constant values of $t_0$ are also associated with a stronger clustering signal,
since this also effectively increases the quasar lifetime $t_q$.

\begin{figure}[t]
  \centering
  \includegraphics[width=0.45\textwidth]{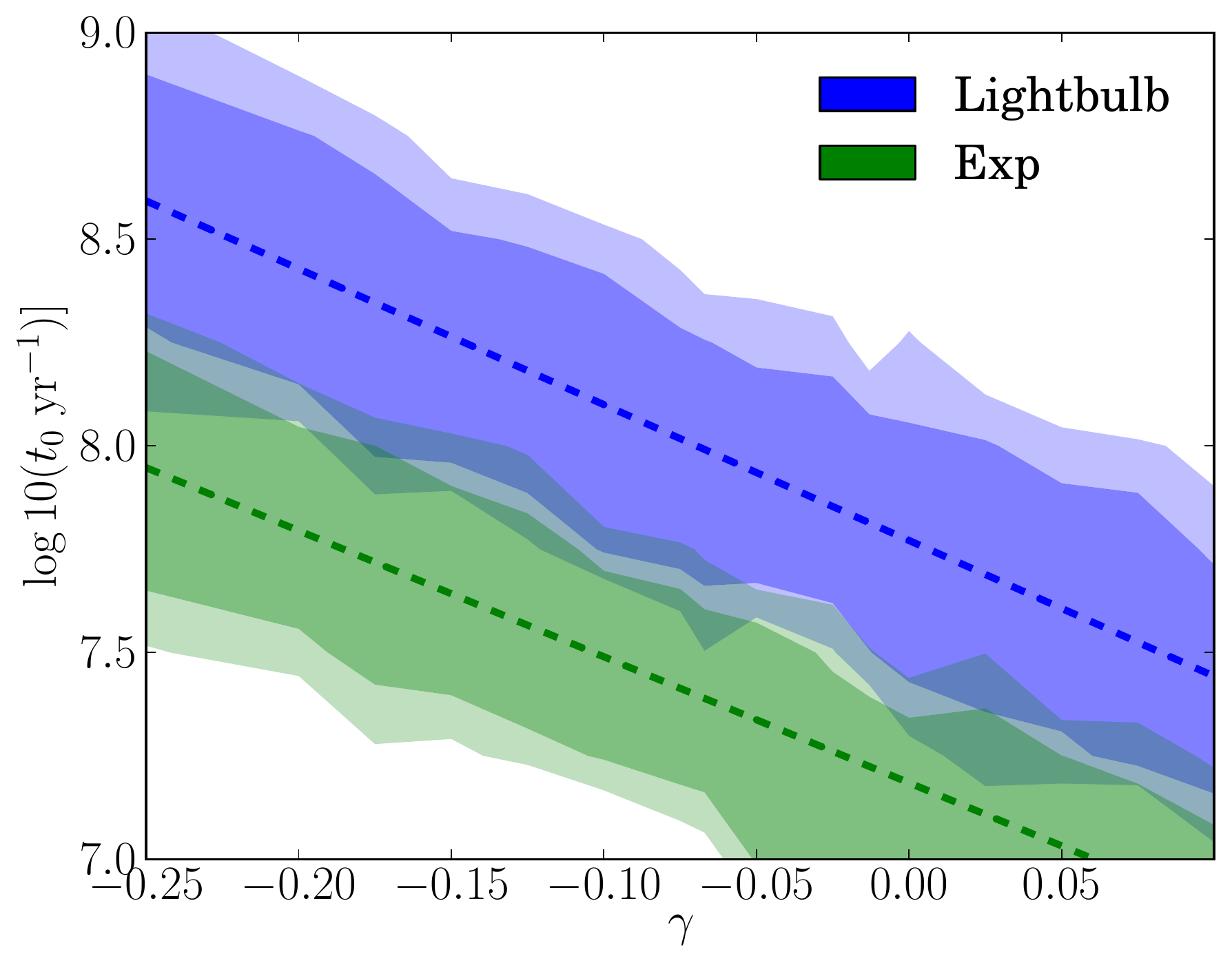}
  \caption{A comparison of the parameter space exploration in terms of the
    parameters $t_0$ and $\gamma$ from Eqn.~(\ref{eqn:tq}). Both the parameter
    space for the lightbulb model (Eqn.~(\ref{eqn:lb})) and exponential model
    (Eqn.~(\ref{eqn:exp})) are shown. The dashed lines represent the best linear
    fits to the data for a particular light curve. The class of models that are
    consistent with the BOSS measurements at $1\sigma$ and $2\sigma$ correspond
    to the darkly and lightly shaded regions. In general, we find that for the
    exponential model, shorter lifetimes are preferred (smaller values of $t_0$
    for the same $\gamma$). Since we abundance match against the quasar's peak
    luminosity, and the quasar spends comparatively little time at or near the
    peak luminosity, we effectively increase the clustering signal for lower
    luminosity quasars.}
  \label{fig:pspace}
\end{figure}

\subsection{Characteristic luminosity and lifetime}
\label{sec:teff}

\begin{deluxetable}{ccccc}
\tablecaption{A List of the Best-fit Parameters for Our Quasar Model as a Function of Redshift. \label{table:teff}}
\tablewidth{0.5\textwidth}
\tablehead{\colhead{Redshift Selection} & \colhead{$z_\mathrm{eff}$} & \colhead{Light Curve} & \colhead{$L_\mathrm{eff}^*$\tablenotemark{a,b}} & \colhead{$t_\mathrm{eff}^*$\tablenotemark{c}}}
\startdata
High-$z$\tablenotemark{d} & 2.51 & Lightbulb & $12.92$ & 7.62 \\
& & Exp & $12.40$ & 7.14 \\[0.5em]
Fiducial & 2.39 & Lightbulb & $13.29$ & 7.77 \\
& & Exp & $13.05$ & 7.18 \\[0.5em]
Low-$z$ & 2.28 & Lightbulb & $13.17$ & 7.84 \\
& & Exp & $13.15$ & 7.29 \\
\enddata
\tablenotetext{a}{$L_\mathrm{eff}$ and $t_\mathrm{eff}$ as defined in Eqns.~(\ref{eqn:teff}-\ref{eqn:Leff}).}
\tablenotetext{b}{$L_\mathrm{eff}^* = \log_{10}(L_\mathrm{eff}/L_\odot)$}
\tablenotetext{c}{$t_\mathrm{eff}^* = \log_{10}(t_\mathrm{eff}/\mathrm{yr})$}
\tablenotetext{d}{The high-$z$ and low-$z$ samples examine the evolution of these parameters with redshift. See Appendix~\ref{sec:appendixb} for further discussion.}
\end{deluxetable}

Figure~\ref{fig:pspace} shows the $\chi^2$ values in the two-dimensional
parameter space $t_0$ and $\gamma$, as defined by Eqn.~(\ref{eqn:tq}), for the
different light curves. The region of good agreement between the BOSS
measurements and our models takes on a linear relationship between
$\log_{10}(t_0)$ and $\gamma$. Such a relationship can be parameterized as:
\begin{equation}
\log_{10}(t_0/\mathrm{yr}) = \log_{10}(t_\mathrm{eff}/\mathrm{yr}) + L_0 \gamma.
\label{eqn:teff}
\end{equation}
The parameters $t_\mathrm{eff}$ and $L_0$ can be thought of as a characteristic
timescale and a characteristic luminosity, respectively. From the functional
form of our power-law for quasar lifetime in Eqn.~(\ref{eqn:tq}), $L_0$ can be
interpreted as changing the normalization luminosity. This is the luminosity at
which all models have the same lifetime, regardless of the value of $\gamma$. In
other words, the characteristic luminosity of the power law becomes:
\begin{equation}
\log_{10}(L_\mathrm{eff}/L_\odot) = 10 - L_0,
\label{eqn:Leff}
\end{equation}
where $L_0$ is defined in Eqn.~(\ref{eqn:teff}). The parameter $t_\mathrm{eff}$
is the characteristic time because all models have this same lifetime at the
luminosity $L_\mathrm{eff}$.

For the lightbulb model, the best-fit values are
$\log_{10}(t_\mathrm{eff}/\mathrm{yr})=7.76$ and
$\log_{10}(L_\mathrm{eff}/L_\odot) = 13.29$. (See Table~\ref{table:teff} for
evolution of these parameters with redshift.)  The characteristic luminosity
inferred from this value is $L_\mathrm{eff} = 10^{13.29}\ L_\odot$, which has a
corresponding magnitude of $M_i = -27.2$. This value is not surprising, given
that quasars were selected for the clustering measurements near this magnitude
range. More interesting is the value of
$\log_{10}(t_\mathrm{eff}/\mathrm{yr})=7.77$, which gives a characteristic
lifetime of $10^{7.77} = 59$ Myr. This is a quasar lifetime that is slightly
longer than those typically quoted in the literature
\citep{yu_tremaine2002,porciani_etal2004,yu_lu2004,conroy_white2013}, which are
closer to the Salpeter $e$-folding time scale or shorter ($\sim$45 Myr for a
quasar accreting at Eddington luminosity and a mass conversion efficiency of
$\epsilon=0.1$). Although $t_\mathrm{eff}$ is slightly higher than these values,
it is within a factor of 2.

\begin{figure}[t]
  \centering
  \includegraphics[width=0.45\textwidth]{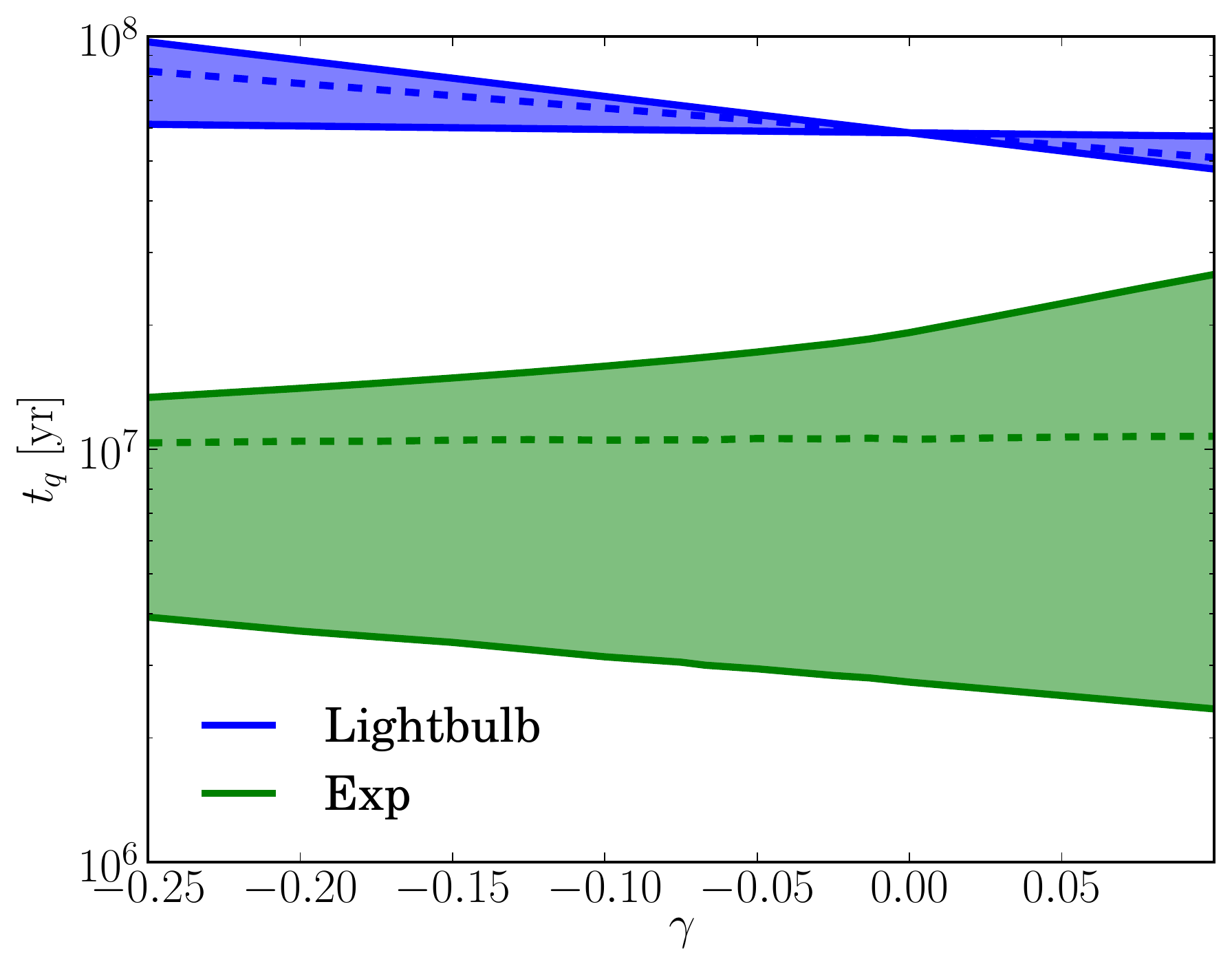}
  \caption{The range of quasar lifetimes for models with the best-fit values of
    $t_0$ as a function of $\gamma$, based on the linear relationship extracted
    in Eqn.~(\ref{eqn:teff}). The solid regions show the span in quasar
    lifetimes, while the dashed line shows the median value within a given
    model. Note that the median value is fairly constant across all quasar
    lifetimes. Thus we are able to characterize a quasar model reasonably well
    using the characteristic lifetime. The comparative large spread in quasar
    lifetime in the exponential case is due to our method of selecting $t_q$,
    rather than reflecting a truly large spread in the data. See the text for
    additional details.}
  \label{fig:lifetime}
\end{figure}

In the exponential model, the best-fit values for $t_\mathrm{eff}$ and
$L_\mathrm{eff}$ defined in Eqn~(\ref{eqn:teff}) are
$\log_{10}(t_\mathrm{eff}/\mathrm{yr})=7.18$ and
$\log_{10}(L_\mathrm{eff}/L_\odot)=13.05$. This luminosity implies a slightly
dimmer characteristic luminosity ($M_i = -26.6$). As discussed in
Sec.~\ref{sec:am}, there is not a single $\tau$ for all quasars for a given
value of $t_\mathrm{eff}$: $L\approx L^*$ quasars have
$\tau \approx t_\mathrm{eff}$, with brighter quasars having
$\tau > t_\mathrm{eff}$. However, the difference between $\tau$ and
$t_\mathrm{eff}$ does not differ by more than a factor of 2 in either direction,
and so to a good approximation $\tau \sim t_\mathrm{eff}$, especially for the
luminosity range used to match the clustering measurements. Compared to the
lightbulb case, the quasars with an exponential light curve have a shorter
characteristic lifetime of 15.1 Myr. The characteristic lifetime is smaller for
the exponential than in the lightbulb case because quasars do not shut off
entirely after a single lifetime, so the time that a quasar is ``bright enough''
to be included within the luminosity cuts is longer than its lifetime
$t_\mathrm{eff}$. This lifetime is about a third of the Salpeter $e$-folding
time scale, which implies that if quasar light curves are roughly exponential,
the combination of the measured QLF and the clustering measurements favors
quasars that either radiate at luminosities dimmer than their Eddington ratio
($L/L_\mathrm{edd} < 1$), have a mass-conversion efficiency that is less that the
fiducial value ($\epsilon < 0.1$), or both. Unfortunately, since our model does
not track the underlying physics present, we are not able to distinguish between
these two cases.

The reason for the different best-fit values between the two models can be
understood as follows. By construction, we have fixed the lifetime of the
exponential quasars such that their peak luminosity-to-mass ratio is the same as
in the case of the lightbulb for a given choice of $t_0$ and $\gamma$. (See
Sec.~\ref{sec:am} for more details.) However, the mass-to-light ratios for the
two light curves are significantly different. This is due to the fact that the
observed luminosity for an exponential quasar can be much smaller than its peak
luminosity. A particular luminosity range is selected for the clustering
measurements, but the clustering of these quasars is tied to their peak
luminosity rather than the observed one. Thus, quasars will tend to have higher
clustering at a given luminosity in the exponential case compared to the
lightbulb, since they spend comparatively little time at or near their peak
luminosity. This luminosity selection includes quasars with a higher peak
luminosity than the chosen range (and thus a higher clustering signal), so we
must also include quasars that have lower mass hosts to match the average
clustering signal. This means that there is a larger spread in host mass
compared to the lightbulb case. This behavior explains why the characteristic
luminosity is slightly smaller for the exponential model compared to the
lightbulb: there is an increased number of low-luminosity quasars occupying
high-mass hosts.

Figure~\ref{fig:lifetime} shows the range of quasar lifetimes as a function of
model parameter $\gamma$. The quasar lifetime is broadly similar across
different model choices. The exponential model has a lower overall value due to
the effect discussed above, \textit{i.e.}, that quasars from a higher peak
luminosity will be included in the sample, bringing along a higher clustering
signal. Since this is true for nearly all the quasars in the sample, there is an
overall decrease in the selected lifetime of quasars. The large difference in
the span of quasar lifetimes is due to the way that we have defined the quasar
lifetime in the exponential model. As discussed in Sec.~\ref{sec:am}, the
exponential lifetime $\tau$ is selected such that the same relationship between
host mass and quasar peak luminosity exists in the exponential case as in the
lightbulb case. Even in a model where for the lightbulb $t_q$ is independent of
$L$ (\textit{i.e.}, when $\gamma=0$), the exponential model parameter $\tau$
does have luminosity dependence. In general, quasars with luminosities above
$L_*$ will have a lifetime longer than an equivalent luminosity in the lightbulb
case for the same choice of $t_0$ and $\gamma$ in Eqn.~(\ref{eqn:tq}), and those
with low luminosities will have a shorter lifetime. This choice for our model
leads to the spread in lifetimes of a factor of $\sim$5, as seen in the case of
$\gamma=0$. For models in which $\gamma > 0$, there is a widening in the range
of values. This is due to the fact that brighter quasars live longer than dimmer
ones. Since our choice of quasar lifetime already enforces this difference,
these models see an increased effect. Conversely, for $\gamma < 0$, there are
competing effects between brighter quasars being less long-lived due to the
choice of $\gamma$, but still simultaneously living longer than their lightbulb
counterparts due to the choice of $t_q$. The latter effect wins out, and these
quasars end up having a significantly larger spread than in the lightbulb
case. Note that the contours in this figure are smooth compared to
Fig.~\ref{fig:pspace} because these are results lying along the best-fit line,
and the figure shows the range in values rather than a single number
(\textit{i.e.}, the $\chi^2$ value) that fluctuates as a function of position in
$t_0$ and $\gamma$.

\section{Discussion}
\label{sec:discussion}
\subsection{Mass-to-light Ratio}
\label{sec:m2l}
The combination of the QLF and clustering measurements produces an important set
of constraints on the space of potential quasar models. Here we investigate the
implications of these models. One important implication is the mass of a typical
halo for a given quasar luminosity. It is trivial to predict this for the case
of a lightbulb model, but less straightforward for the case of the exponential
model. Here the peak luminosity $L_\mathrm{peak}$ is used to define the
mass-to-light ratio, since this is the quantity used in our abundance matching
approach. This ratio defines a typical mass for quasars, which can be compared
with results of previous analysis (\textit{e.g.},
\citealt{martini_weinberg2001,shen_etal2007,white_etal2012}).

\begin{figure}[t]
  \centering
  \includegraphics[width=0.45\textwidth]{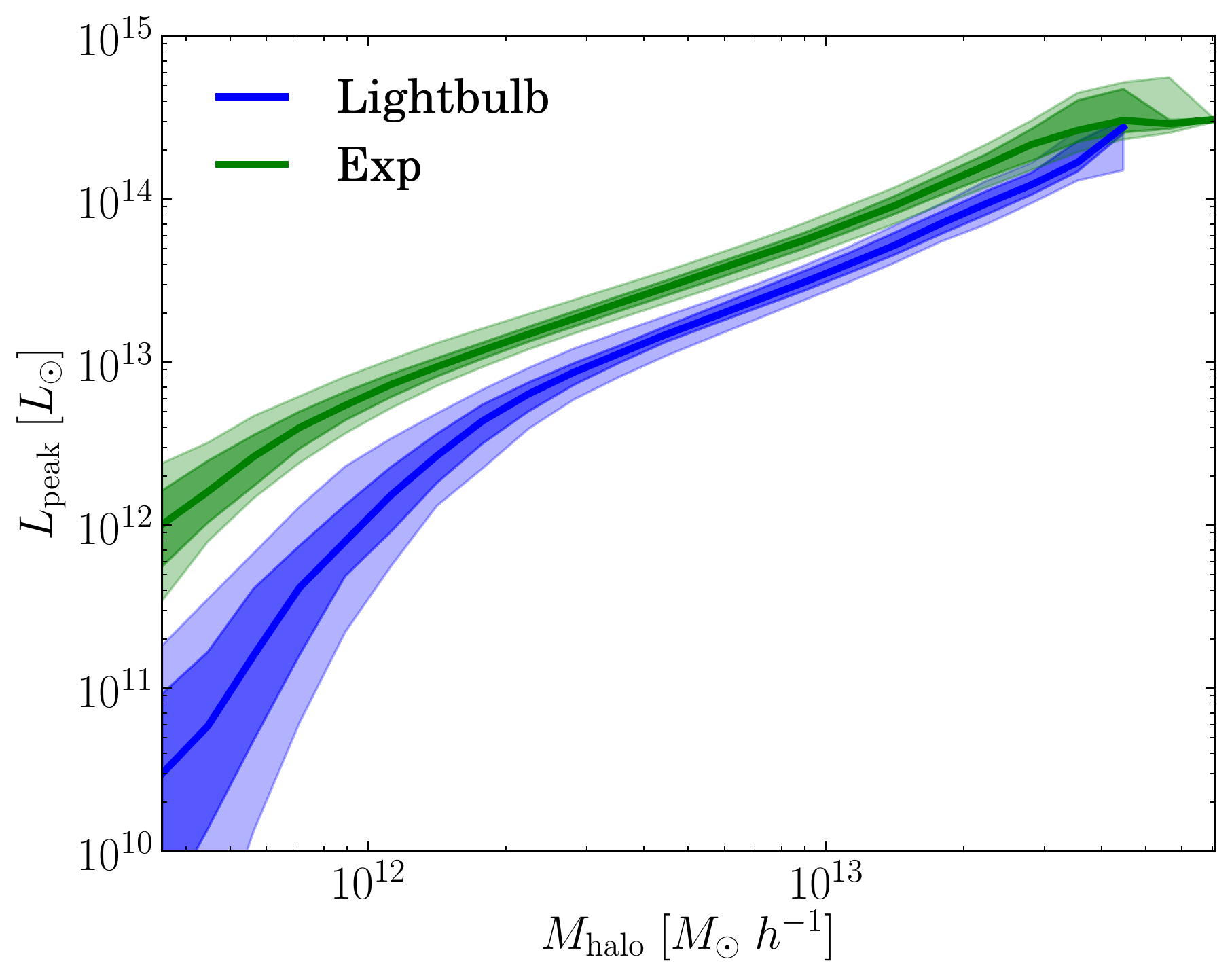}
  \caption{A comparison of the mass-to-light ratio between the different quasar
    models. We have computed this ratio for each model in our parameter space,
    and weighted their contribution by their corresponding $\chi^2$ value. The
    lines show the median value by weight and the shaded regions show
    $\pm 1\sigma$ and $2\sigma$. Note that for a given luminosity, a quasar in
    the exponential model is found in a halo with a smaller mass. This is due to
    the fact that quasars with a peak luminosity significantly greater than the
    observed one are included in the luminosity range selected for the
    clustering measurements. These hosts have a higher clustering signal than
    quasars with a peak luminosity in the luminosity selection. This means we
    must also select lower-mass objects as well. See the text for additional
    details.}
  \label{fig:m2l}
\end{figure}

Figure~\ref{fig:m2l} shows the luminosity of quasars as a function of the mass
of the host halo for the lightbulb and exponential models for all combinations
of $t_0$ and $\gamma$. This plot shows the mass-to-light ratio of the entire
catalog. The weight assigned to the luminosity as a function of mass
$L_\mathrm{peak}(M)$ for a particular model $i$ is given by a $\chi^2$
likelihood. We also find the $\pm$ 1 and 2 $\sigma$ values that enclose 68\% and
95\% of the likelihood.

\begin{figure}[t]
  \centering
  \includegraphics[width=0.45\textwidth]{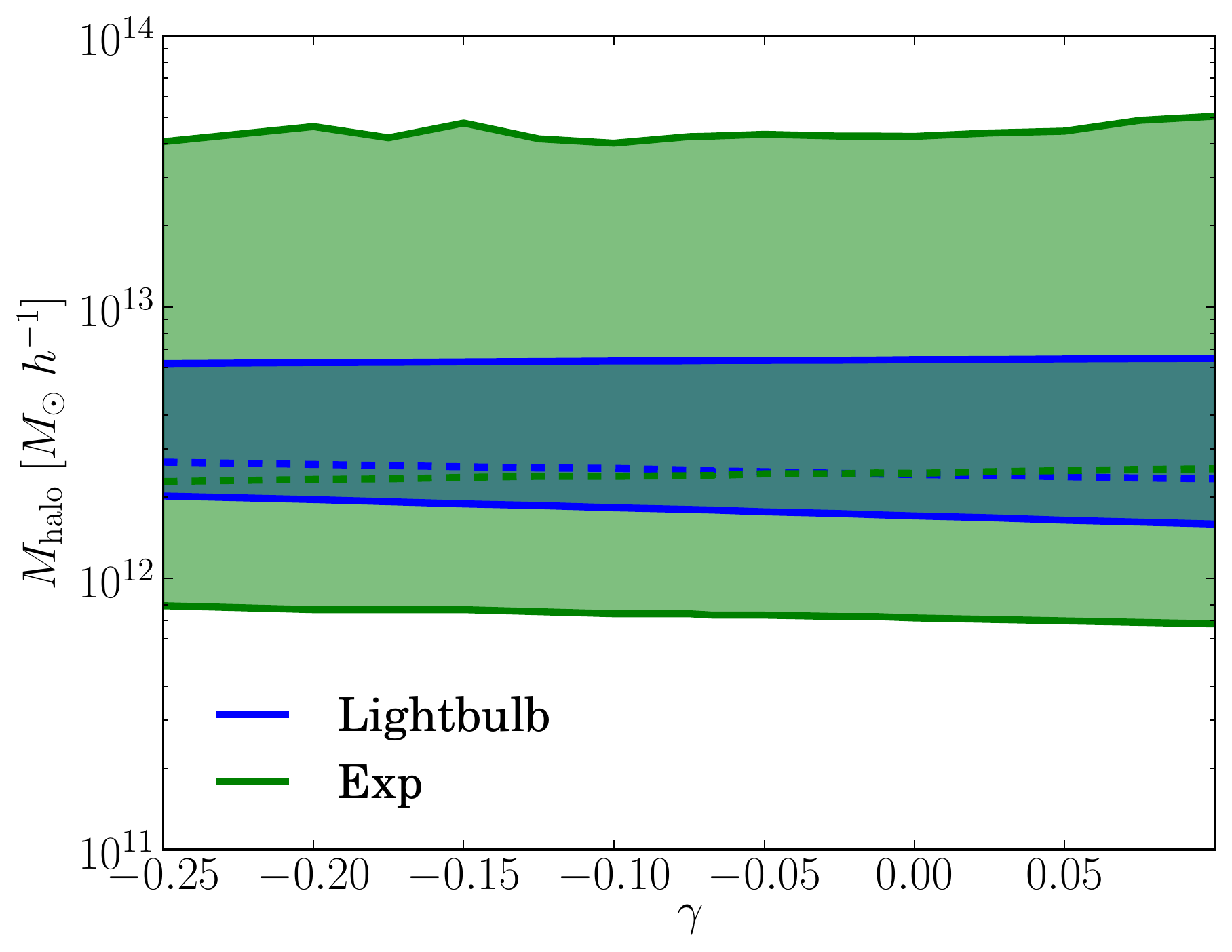}
  \caption{The selected mass range for our different models. For the lightbulb
    case, there is a relatively narrow range in mass. Since the clustering of
    quasars is fixed as a function of luminosity, there is a very tight
    relationship between observed luminosity and underlying host mass. In the
    exponential case, the range is much more extended. Since the mass-to-light
    ratio is fixed to be the same for the peak luminosity, the evolution within
    the model means that there will be quasars with higher peak luminosity (and
    higher-mass hosts) selected by the evolution.}
  \label{fig:host_mass}
\end{figure}

Figure~\ref{fig:host_mass} shows the mass range as a function of the model
parameter $\gamma$. Note that the range is essentially constant with respect to
$\gamma$. By averaging the median mass across all values of $\gamma$, a
characteristic mass for the two models can be defined. This characteristic mass
is $2.5\times 10^{12}$ $h^{-1}M_\odot$ for the lightbulb model, and
$2.3 \times 10^{12}$ $h^{-1}M_\odot$ for the exponential model. These values are
broadly consistent with previous studies of quasar clustering measurements
(\textit{e.g.},
\citealt{porciani_etal2004,croom_etal2005,lidz_etal2006,porciani_norberg2006,white_etal2012}). Since
in all models the same clustering signal of the quasars is being selected, there
is an implicit requirement for the hosts to lie within a certain mass
range. Additionally, the mass range in the exponential case is significantly
larger than that of the lightbulb model. This is again related to the fact that
due to the light curve, quasars with a higher clustering signal are included
within the luminosity sample, and so there must also be lower-mass hosts
included as well to balance the average clustering strength. There is a
significantly larger spread above the median mass than below. The reason for
this asymmetry is due to the difference in number density: since the high-mass
objects are rarer, a comparatively smaller range in low-mass halos is necessary
to make the clustering signal equivalent to the lightbulb case. See
Sec.~\ref{sec:teff} for further discussion.

In the case of the lightbulb model, the halo mass that corresponds to the
selected luminosity range of quasars is relatively tightly constrained. For the
models that agree with the BOSS measurements at $1\sigma$, the average halo mass
ranges from 1.35~$\times$~10$^{12}$ to 4.93~$\times$~10$^{12}$~$h^{-1}M_\odot$
for hosts of quasars within the magnitude cutoff. For the exponential model,
there is a much larger range in halo mass: for the collection of models that
agree at $1\sigma$, the mass ranges between 6.69~$\times$~10$^{12}$ and
5.85~$\times$~10$^{13}$~$h^{-1}M_\odot$, almost an entire order of magnitude
(compared to about half an order of magnitude for the lightbulb model). Also
note that the mass range is much larger than in the lightbulb case. This fact
can be explained by noting that there is evolution in the mass-to-luminosity
ratio during the lifetime of the quasar. Further, the $e$-folding time for these
models is comparatively long, with typical values being $\tau \approx 40$
Myr. This means that there are high-mass hosts included in the sample of quasars
chosen for the clustering measurements whose quasars are not at their peak
luminosity. Since these hosts have a bias larger than the value preferred by the
BOSS measurements, this sample must necessarily include hosts which have a
smaller clustering signal, so that on average, the total bias agrees with BOSS.

There is a systematic shift upward in the mass of the exponential case compared
to the lightbulb. This shift is related to the difference in parameter space
discussed in Sec.~\ref{sec:2pcf}. Due to their exponential change in luminosity,
the quasars are not typically found near their peak luminosities. Thus, even
though by construction the peak quasar luminosity as a function of halo mass is
the same for the two models, the effective luminosity for a given mass is
reduced in the exponential case due to the light curve evolution. In other
words, quasars of the same luminosity in the two different models are found in
more massive hosts in the exponential case. This leads to a systematic shift in
the preferred mass range for clustering measurements. To sum up: the increased
spread in halo host mass for the exponential model compared to the lightbulb is
due to inclusion of highly biased hosts in the measurement being balanced out by
lower-mass ones, and the systematic shift toward higher mass is due to the
effective increase in the mass-to-luminosity ratio related to evolution of
quasar luminosity.

\subsection{Mass function and duty cycle of halo hosts}
\label{sec:mass_func}

\begin{figure}[t]
  \centering
  \includegraphics[width=0.45\textwidth]{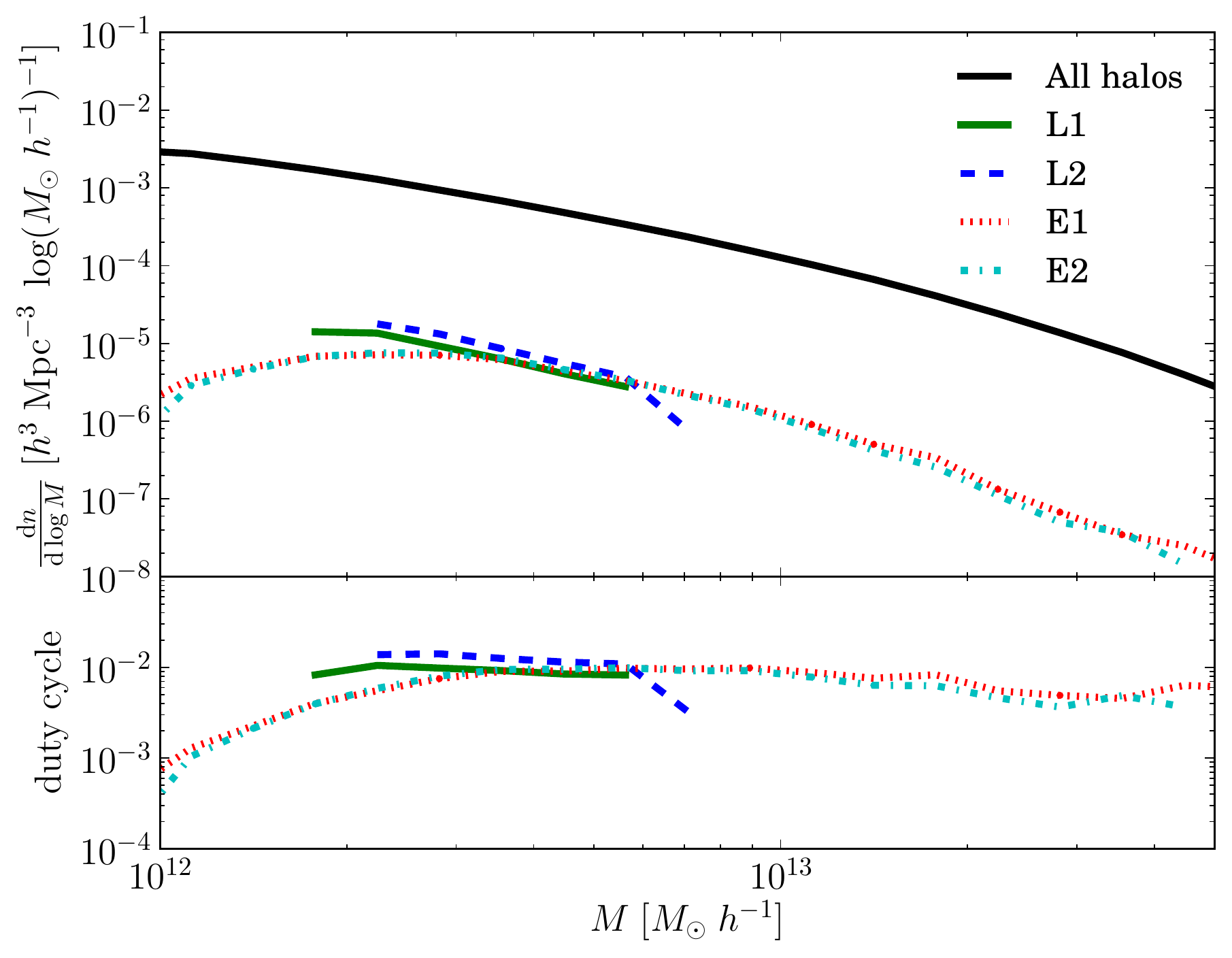}
  \caption{The halo mass function for halos hosting quasars versus the total
    halo population for certain models as defined in
    Table~\ref{table:models}. In the case of the lightbulb model, there is a
    roughly constant value of $\sim$1\% of halos hosting quasars as a function
    of halo mass. For the exponential, there is a constant active fraction at
    the high-mass end, but then the fraction falls off below the median halo
    mass. This is again related to the fact that there are more halos included
    at low mass in order to reproduce the average clustering signal.}
  \label{fig:mass_func}
\end{figure}

To observe the effect that different points in parameter space have on the halo
host properties, the mass function of halos hosting an active quasar has been
calculated for the fiducial redshift selection. (For the high- and low-redshift
selections, see Appendix~\ref{sec:appendixb}.) Figure~\ref{fig:mass_func} shows
the total halo mass function as well as the mass function of halos hosting
quasars within the luminosity range $-25 \geq M_i \geq -27$. From this analysis,
the duty cycle of halo hosts can be extracted, \textit{i.e.}, the fraction of
active halos divided by the total number of halos. As discussed in
Sec.~\ref{sec:light_curve}, in the lightbulb model the duty cycle can be
directly related to the quasar lifetime at that luminosity. However, here the
duty cycle is defined simply as the active fraction of halos.

Figure~\ref{fig:mass_func} compares the case of the lightbulb and the
exponential light curves. When the halo mass function of active halos is
examined in the two different cases, it can be seen that the mass range of hosts
spanned by an individual model is quite different. In the lightbulb case, there
is a very small range in host mass compared to the exponential case. This
difference can be explained in terms of which hosts are included in the
clustering measurements. In the lightbulb case, since the luminosity is constant
as a function of quasar lifetime, the only evolution in the relationship between
mass and luminosity comes from mass accretion, which makes up a small fraction
of total halo mass over the time scales for which quasars are active. As such,
with an essentially static relationship, there is a very strong correlation of
mass to light. For a specific model, there is only about a factor of 2 in halo
mass included for the quasars in the selected magnitude range. When looking at
the duty cycle of quasar hosts, one can see that the fraction is typically
0.5-1\%, with little evolution with mass within a model.

Conversely, in the exponential model, there is evolution for individual halos in
the mass-to-light relationship. Most importantly, this implies that massive
halos will be included when selecting quasars at a specific luminosity. Since
they are more highly clustered (and more biased), smaller, less biased halos
must also be included in order to create an average bias consistent with the
BOSS measurements. This has the effect of extending the mass range of halos
included in the mass function. Note that within a single model, there is a much
larger span in halo mass: in some cases, the span is more than an order of
magnitude in halo mass. Additionally, the duty cycle is comparable in magnitude
to the lightbulb case, though slightly smaller: the ratio of active halos to
total halos ranges from 0.05-1\%. There also seems to be a trend in the
evolution of the duty cycle: there is a central ``typical mass'' for a given
model, and the duty cycle decreases in both directions. A similar trend was
found by \citet{white_etal2012}.

Note that one result of this measurement is the fact that the mass range of host
halos is significantly more extended in the exponential case than the lightbulb
case. Thus, one way to break the degeneracy between the lightbulb and
exponential models would be to measure the mass range of underlying host halos,
perhaps through using gravitational lensing to independently find the mass of
the dark matter halo \citep{courbin_etal2012}. If the range of masses for quasar
hosts is extended, then there would be observational evidence favoring an
exponential model (or a model with evolution in the quasar light curve) as
opposed to the lightbulb model.

\section{Predictions for Helium Reionization}
\label{sec:reion}

\begin{figure*}[t]
    \begin{minipage}[t]{0.47\hsize}
    \centering{\small QLF:}
  \end{minipage}
  \begin{minipage}[t]{0.47\hsize}
    \centering{\small $L_{912}$ normalization:}
  \end{minipage}
  \centering{
    \resizebox{0.47\hsize}{!}{\includegraphics{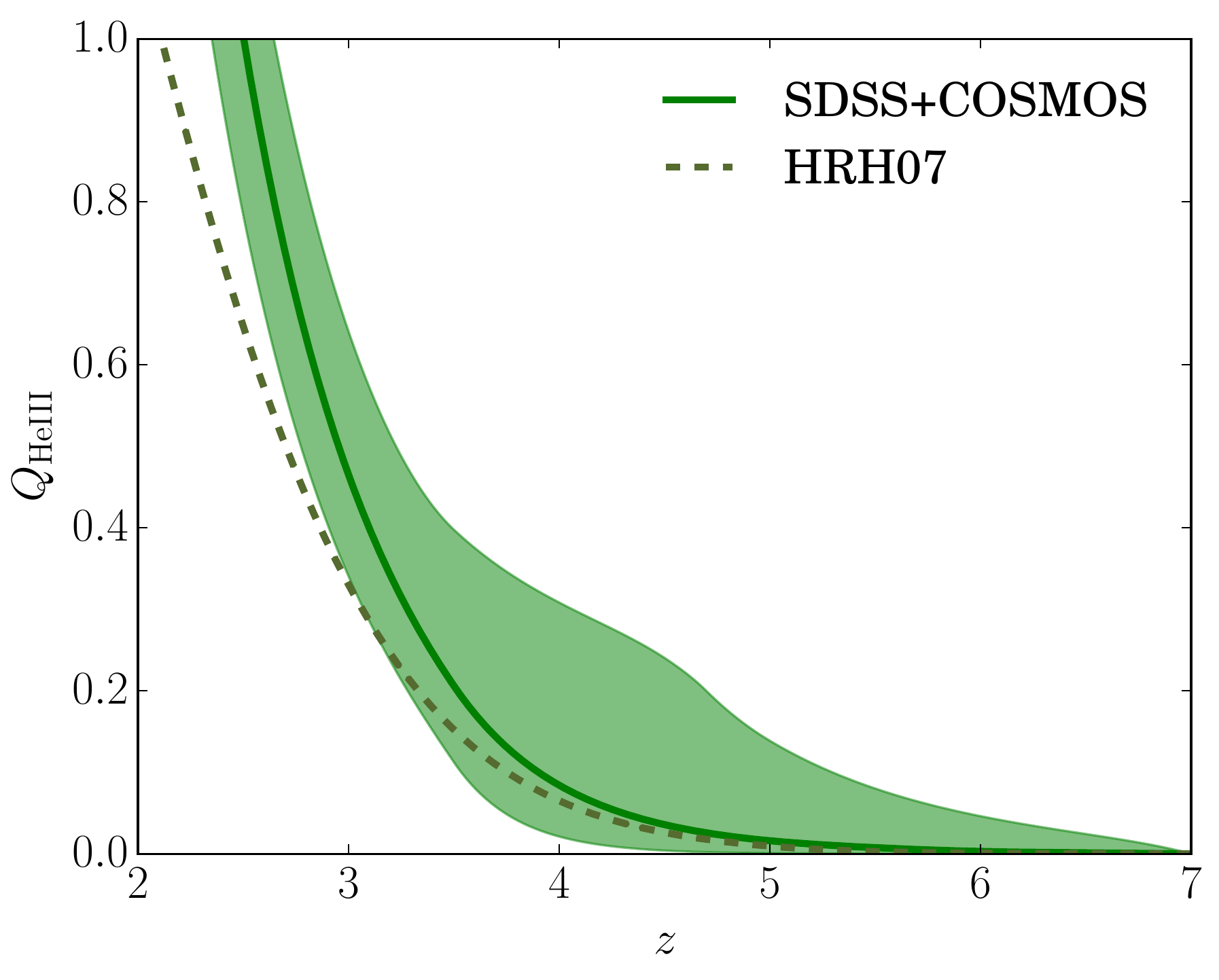}}%
    \hspace{2pt}%
    \resizebox{0.47\hsize}{!}{\includegraphics{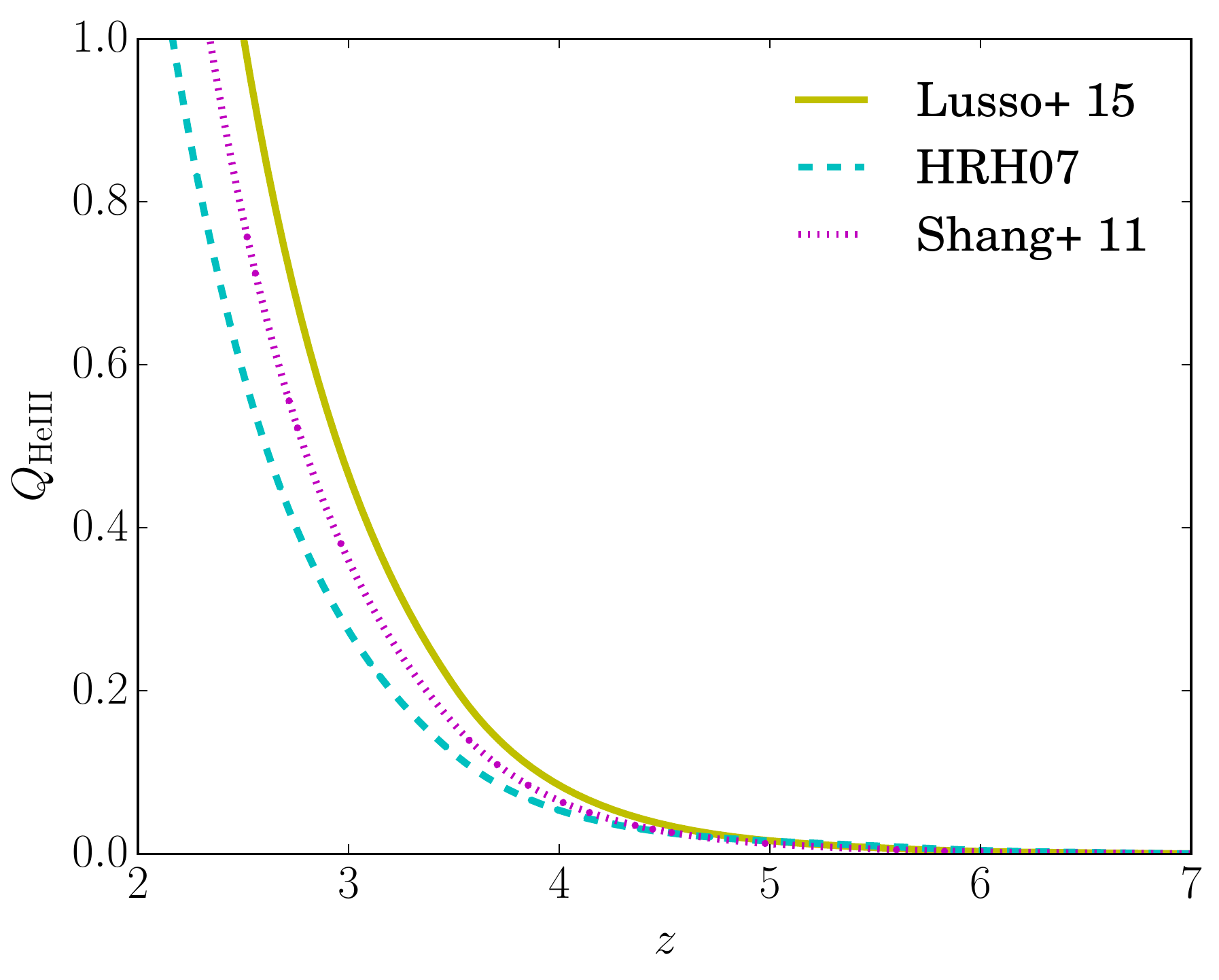}} }\\
  \begin{minipage}[t]{0.47\hsize}
    \centering{\small SED spectral index:}
  \end{minipage}
  \begin{minipage}[t]{0.47\hsize}
    \centering{\small Clumping factor:}
  \end{minipage}
  \centering{
    \resizebox{0.47\hsize}{!}{\includegraphics{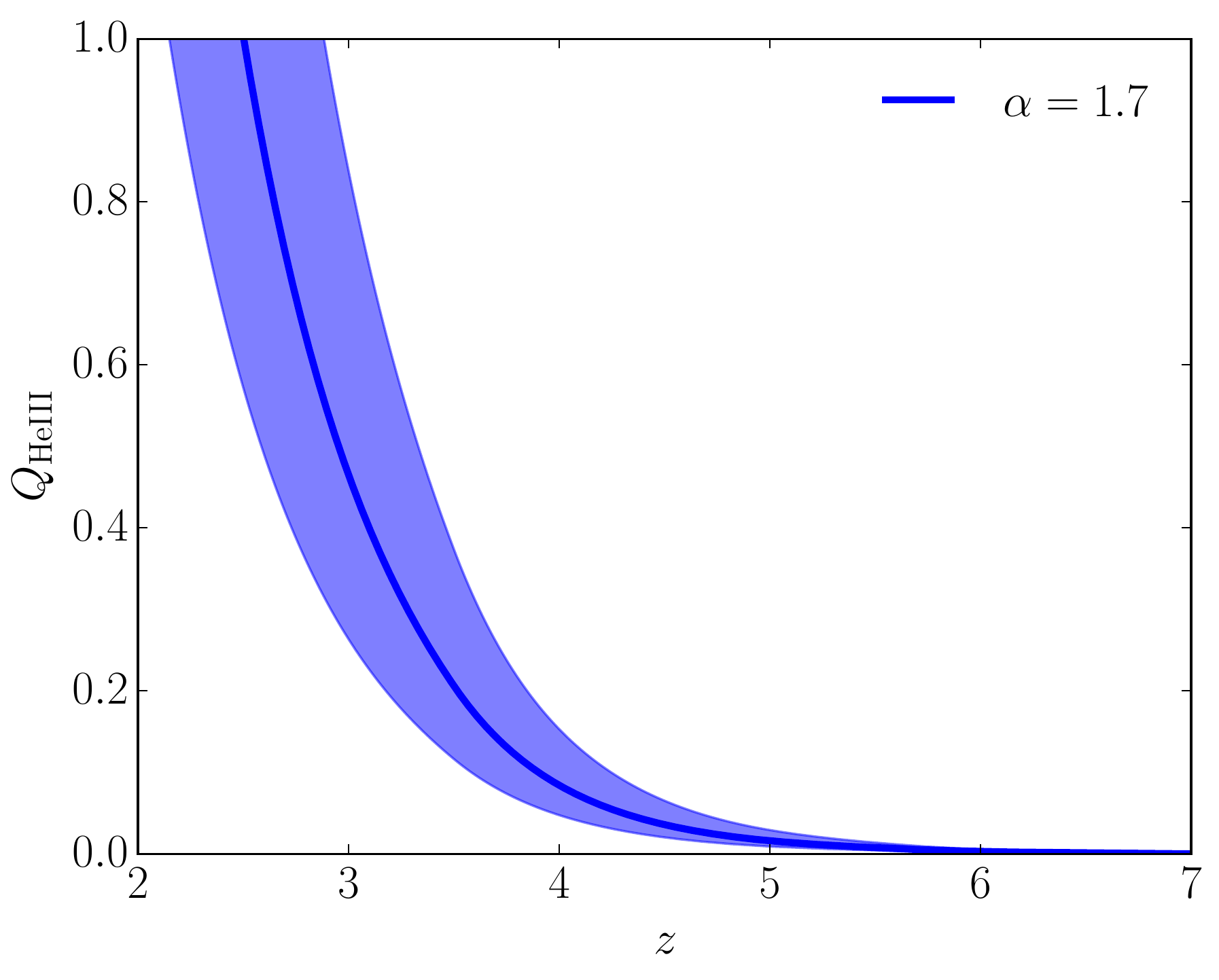}}%
    \hspace{2pt}%
    \resizebox{0.47\hsize}{!}{\includegraphics{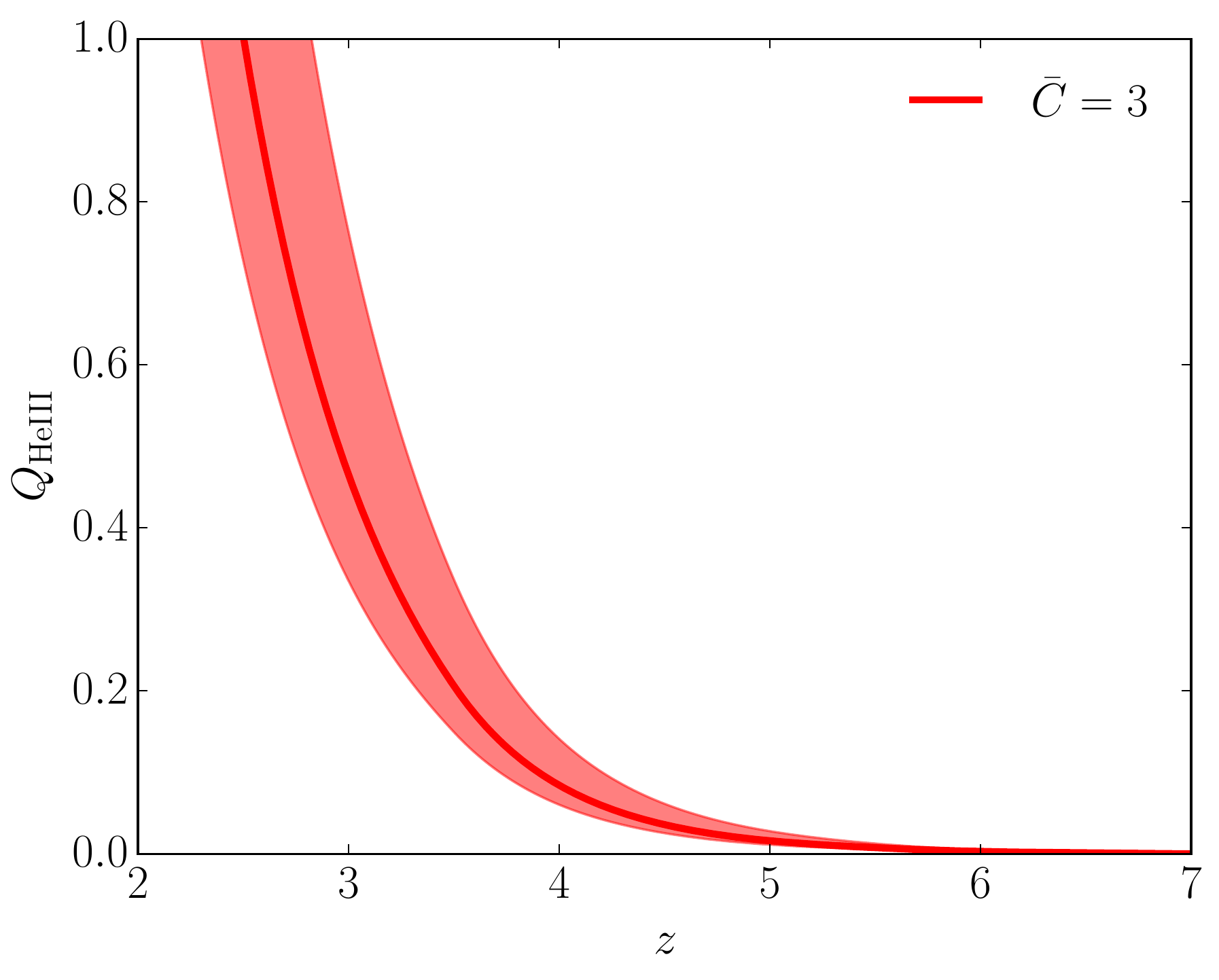}} }\\
  \caption{The volume-filling fraction $Q_i$ of doubly ionized helium defined in
    Eqn.~(\ref{eqn:volfrac}). In each plot, we show the fiducial values we have
    for $Q_i$ as a function of redshift, which has the parameters $\alpha=1.7$,
    $\bar{C}=3$, and normalizing the luminosity at 912 \AA\ following
    \citet{lusso_etal2015}. This leads to a redshift of reionization of
    $z \sim 2.5$, comparable to the redshift of $z \sim 2.7$ suggested by
    observation of the helium Lyman-$\alpha$ forest. We show the change in $Q_i$
    as a function of varying these parameters. Top left: we compare the
    difference between using the composite QLF of SDSS+COSMOS (see
    Sec.~\ref{sec:qlf} for more details) and the one in
    \citet{hopkins_etal2007}. The shaded region reflects differences in
    ionization level due to jointly varying the parameters over the ranges
    specified in Table~\ref{table:parameters}. Top right: we change the UV SED
    of the quasar, which affects the normalization at 912 \AA. In addition to
    the SEDs from \citet{lusso_etal2015} and HRH07, we show the radio-quiet
    template from \citet{shang_etal2011}. Bottom left: we allow the EUV SED
    spectral index for $\lambda < 912$ \AA\ to vary from
    $1.4 \leq \alpha \leq 2.0$. Bottom right: we vary the clumping factor of the
    IGM, from $1 \leq \bar{C} \leq 5$. See the text for additional details.}
   \label{fig:volfrac}
\end{figure*}

One very important prediction that we can make from our quasar catalog is the
redshift of helium reionization. In order to understand in detail the
implications for helium reionization, we need to run full hydrodynamic plus
radiative transfer numerical simulations. However, we can perform a
semi-analytic calculation in order to find a rough estimate of the redshift of
reionization by computing the fraction $Q_i$ of the universe's volume that has
been reionized (also called the volume-filling fraction), where $Q_i = 1$
represents a totally reionized universe (\textit{e.g.},
\citealt{madau_etal1999,furlanetto_oh2008}):
\begin{equation}
\dv{Q_i}{t} = \int \dd{L} \frac{\dot{N}_\gamma}{\bar{n}_\mathrm{He}} \dv{\phi}{L} - \bar{C}\alpha_A \bar{n}_e Q_i,
\label{eqn:volfrac}
\end{equation}
where $\bar{n}_\mathrm{He}$ is the number density of neutral helium, $\bar{n}_e$
is the number density of electrons, $\dot{N}_\gamma$ is the production rate of
ionizing photons for an individual quasar, $\alpha_A(T)$ is the recombination
coefficient, and $\bar{C} \equiv \ev{n_e^2}/\ev{n_e}^2$ is the clumping factor
of the ionized IGM. The minimum luminosity of the integral decreases as redshift
decreases, in keeping with modeling and observations
\citep{richardson_etal2012,shen_kelly2012,cen_safarzadeh2015,sijacki_etal2015}. The
clumping factor measures the effective distribution of gas inside the scale of
volume being averaged (or resolution in the case of simulations). Note that
these calculations assume a primordial helium mass fraction of
$Y_\mathrm{He} = 0.24$. Following the arguments in the appendix of
\citet{kaurov_gnedin2014}, we choose the case A recombination coefficient, which
assumes that photons emitted from recombination are not reabsorbed by a neutral
atom, increasing the recombination rate. \footnote{Although the arguments
  presented in the cited work are in the context of hydrogen reionization, the
  same arguments can be applied equally well to helium
  reionization. Essentially, the authors argue that the photons redshift out of
  resonance with the thermally broadened spectral line before they encounter the
  edge of the ionized region or a Lyman-limit system. Although the ionization
  fraction of helium might be slightly lower inside an ``ionized region'' than a
  comparable hydrogen one, the difference is not significant enough to change
  the overall conclusion.} It is assumed that initially, all of the hydrogen in
the IGM has been ionized, and all of the helium is singly ionized. To compute
the photoionization rate of an individual quasar $\dot{N}_\gamma$, the SED of
\citet{lusso_etal2015} is used to convert the specific luminosity at 2500 \AA\
to that at 912 \AA. It is then assumed that quasars have an SED that follows a
power law $L_\nu(\nu) \propto \nu^{-\alpha}$ for values of $\lambda < 912$
\AA. The fiducial value chosen is $\alpha = 1.7$, also based on observations of
the rest-frame UV spectra of quasars from \citet{lusso_etal2015}. This
calculation includes all photons with frequencies in the range
54.4~eV~$\leq$~$h\nu$~$\leq$~1~keV. Photons above this energy have a mean free
path of helium ionization comparable to the Hubble distance.

Although the two different quasar light curves explored above have different
individual properties, both are constrained by the global properties fixed by
the QLF. We find that if instead of the statistical calculation outlined above,
we use the number of ionizing photons computed directly from the quasar
catalogs, the result differs only by a few percent. Therefore, it is much more
straightforward to use Eqn.~(\ref{eqn:volfrac}). This approach also permits the
use of other QLFs in the calculation, so it is possible to explore what effect
this has on the results.

Figure~\ref{fig:volfrac} shows the ionization fraction as a function of redshift
computed from Eqn.~(\ref{eqn:volfrac}). In the first panel, there is a
comparison of the choice of QLF used in the calculation. Included are the QLF
used in the main body of this work, the composite QLF composed of the ones from
R13, M12, and M13 (the SDSS+COSMOS) as explained in Sec.~\ref{sec:qlf}, and the
QLF from \citet{hopkins_etal2007} (hereafter referred to as HRH07). All other
calculations presented use the composite QLF, but then change various other
parameters. Note that for the prediction of reionization time using HRH07, both
the QLF and the SED are different from the fiducial comparison case. In the
figure, the shaded region shows the range of predicted values for the
volume-filling fraction $Q_i$ at a given redshift $z$ by jointly varying the
parameters of the QLF over the range specified by
Table~\ref{table:parameters}. Interestingly, the late-time ionization level is
less sensitive to the variation in parameters at early redshift, due to the
interplay between the source and recombination terms present in
Eqn.~(\ref{eqn:volfrac}). At redshifts $z \leq 3.5$, the source term becomes the
same for all histories, since the QLF transitions to that of Ross et
al. Further, the recombination rate is proportional to the ionized fraction, so
histories that had higher ionization levels at $z \geq 3.5$ will have higher
levels of recombination. Since the recombination time is much shorter than the
total timescale of the reionization calculation, all histories converge on a
similar redshift of total reionization ($Q_i=1$). Nevertheless, the variation in
ionization fraction at early times can have important implications on the
topology of ionized regions and the thermal history of the IGM, so such
differences may in principle be detectable.

In the second panel of the plot, the specific luminosity of individual objects
at 912 \AA\ $L_{912}$ is varied. One way to achieve this variation is the change
the UV SED template used for quasars. Once the specific luminosity $L_{2500}$ is
calculated from the observed magnitude according to Eqn.~(\ref{eqn:L2500}), the
quasar SED can be used to find $L_{912}$. In the fiducial approach, we use the
SED template from \citet{lusso_etal2015}, which assumes a UV spectral index of
$\alpha = 0.61$ for $2500\ \mathrm{\AA} \geq \lambda \geq 912\ \mathrm{\AA}$. An
alternative choice for an SED is one from \citet{shang_etal2011}, which provides
a composite quasar SED template by combining observations in different frequency
ranges to create a single spectrum. \citet{shang_etal2011} divide the sample
into radio-loud and radio-quiet quasars. However, radio-quiet quasars compose
$\sim 90\%$ of high-redshift quasars found in the SDSS
\citep{shen_etal2009}. Thus, we only include the results of the calculation
using the radio-quiet template. This template provides the relative specific
luminosity at each frequency, and so can be used to convert $L_{2500}$ to
$L_{912}$. The effective spectral index for this wavelength range for the
radio-quiet quasar template is $\alpha = 0.867$. In addition, we show the impact
of using the SED from HRH07 (with the QLF from Sec.~\ref{sec:qlf}). Note that
the SED from HRH07 is outdated, and used only as a point of comparison. More
recent studies (\textit{e.g.}, \citealt{stevans_etal2014,lusso_etal2015}) are
largely inconsistent with the SED of HRH07, and so it is presented here merely
to emphasize the importance that using the proper SED has on helium
reionization. Given this same specific luminosity $L_{2500}$, the predicted
value of $L_{912}$ from the SED of \citet{lusso_etal2015} is higher than that of
HRH07 by about a factor of 1.7, leading to the earlier reionization time. The
second panel of the plot includes these to demonstrate the difference from using
different quasar templates.

In the third panel of the plot, the spectral indices are varied, ranging from
$1.4 \leq \alpha \leq 2.0$. Recent measurements from \citet{lusso_etal2015}
suggest that at high redshift and bright magnitudes, the spectral index has a
value of $\alpha = 1.7 \pm 0.6$. This is slightly softer than the average value
of $\alpha = 1.6$ from \citet{telfer_etal2002}. In order to explore some of the
implications of changing the spectral index, we vary its value as indicated.

The final panel explores a range of clumping values, from
$1 \leq \bar{C} \leq 5$. The precise value for the clumping factor for helium
reionization is very uncertain, as most studies on the clumping factor are
related to hydrogen reionization (see, \textit{e.g.},
\citealt{raicevic_theuns2011,kaurov_gnedin2014}). In \citet{furlanetto_oh2008},
the authors explored clumping factors of $0 \leq \bar{C} \leq 3$. More recent
results from numerical simulations were calculated by
\citet{jeeson-daniel_etal2014}, who found that the clumping factor of helium
ranges from $3 \leq \bar{C} \leq 8$ for the redshift range of interest,
depending on the ionization level of the helium gas.

In Figure~\ref{fig:volfrac}, each panel shows the fiducial evolution of $Q_i$,
which is characterized by the values of $\alpha=1.7$, $\bar{C}=3$, the SED of
\citet{lusso_etal2015}, and the composite SDSS+COSMOS QLF. In this situation,
the redshift of reionization (\textit{i.e.}, when $Q_i=1$) is $z \sim 2.5$. This
value is comparable to, though slightly later than, the redshift suggested by
recent observations of $z \sim 2.7$
\citep{dixon_furlanetto2009,worseck_etal2011}. However, a smaller
volume-averaged clumping factor $\bar{C}$ or a larger amplitude in either the
measured QLF or the specific luminosity $L_{912}$ could give an earlier redshift
of reionization. Specifically, assuming the fiducial model, changing the
clumping factor to $\bar{C} = 1.7$ would give $z \sim 2.7$ as the redshift of
reionization. It should be noted that this calculation is not wholly accurate
for reionization, since it assumes a single clumping factor for the entire IGM,
which is almost certainly not accurate for helium reionization, due to its very
inhomogeneous nature. Furthermore, this calculation does not include secondary
ionizations from energetic electrons (\textit{e.g.},
\citealt{shull1979,furlanetto_stoever2010}), though these interactions are
likely unimportant for helium reionization \citep{mcquinn_etal2009}.

When comparing to the results of \citet{furlanetto_oh2008}, we notice that the
authors' value for the redshift of reionization is significantly earlier than
the one that we have found. This is largely due to a different QLF used, as well
as a different method for calculating a quasar's EUV SED. The referenced paper
uses the QLF from HRH07, and assumes an SED that gives more EUV radiation. This
luminosity function has a significantly larger amplitude compared to the results
from R13, up to an order of magnitude larger for low-luminosity quasars at high
redshift. (See Fig.~16 of \citealt{ross_etal2013}.) Thus, accurate measurements
and a proper understanding of the systematics of the high-redshift QLF, as well
as the accompanying quasar SED, are essential for a proper treatment of helium
reionization.

\section{Conclusion}
\label{sec:summary}
We have provided a technique for populating dark matter halos with quasars that
matches a QLF by construction for various light curve models of quasars. By
using the triggering rate of \citet{hopkins_etal2006} with the technique of
abundance matching, we are able to match the observed QLF of SDSS Data Release 9
(DR9) (R13), COSMOS (M12), and high-redshift SDSS data (M13). After applying
this method to dark matter halo catalogs generated from $N$-body simulations, we
have constrained a class of quasar models that reproduce the clustering
amplitude measured from the two-point auto-correlation function of the BOSS
survey \citep{white_etal2012} at a redshift of $z=2.39$. The characteristic mass
of the quasar hosts is $2.5 \times 10^{12}$ $h^{-1}M_\odot$ for the lightbulb
model and $2.3 \times 10^{12}\ h^{-1}M_\odot$ for the exponential model. The
effective lifetime as defined in Eqn.~(\ref{eqn:teff}) of quasars is
$t_\mathrm{eff} = 59$ Myr for the lightbulb model of quasars and
$t_\mathrm{eff} = 15$ Myr for the symmetric exponential model.

One of the limitations of this approach is that we have constrained the class of
quasar models using a comparatively narrow span in quasar luminosity. By
matching the bias of quasars with a different magnitude range, we would have a
different effective luminosity range for the bias calculation. This would lead
to a different slope in the parameter $L_\mathrm{eff}$, which would allow us to
break the degeneracy observed in Fig.~\ref{fig:pspace}. Having the ability to
break the sample down into different luminosity intervals would allow us to make
tighter constraints on the class of allowed models.

In future work, we plan to use the quasar models explored here as sources of
ionizing photons for studying helium reionization using simulations containing
hydrodynamics and radiative transfer. These types of simulations will allow us
to accurately capture important physical characteristics related to the
IGM. Specifically, we are interested in capturing the thermal history of the IGM
as it relates to observations. In upcoming simulations, we plan to compute the
IGM equation of state and produce synthetic Lyman-$\alpha$ forest fluxes. This
will allow us to tap into the wealth of observations available for the
Lyman-$\alpha$ forest, such as those currently available from BOSS
(\textit{e.g.}, \citealt{lee_etal2013}), and from upcoming future surveys such
as DESI.

\acknowledgments{We thank the referee for many constructive comments, which
  helped improved the final manuscript. We would like to thank Ross O'Connell,
  Renyue Cen, and Ying Zu for helpful discussions. We thank Simeon Bird and
  Michelle Ntampaka for useful comments on drafts of this work. HT is supported
  in part by NASA grant NNX14AB57G and NSF grant AST 1312724.}

\bibliography{mybib}
\bibliographystyle{apj}

\begin{appendix}

\section{Fitting the parameters of the QLF}
\label{sec:appendixa}
In order to construct a QLF informed by the observations at all redshifts
relevant to helium reionization, we have combined the measurements of R13, M12,
and M13. We will now briefly summarize the relevant findings of each paper. In
all three results, the QLF is parameterized as a double-power law, according to
Eqn.~(\ref{eqn:qlf}). R13 uses quasars identified from SDSS-III DR9, and
provides a LEDE model in which the base-10 logarithm of the QLF normalization,
$\log_{10}\phi^*$, and the break magnitude $M^*$, evolve linearly with redshift,
as parameterized in Equations~(\ref{eqn:ledephi}) and (\ref{eqn:ledem}). The
parameters $\alpha$ and $\beta$ are fixed as a function of redshift. Nominally,
the LEDE fit is valid over the redshift range $2.2 \leq z \leq 3.5$. M12 uses
data from the COSMOS survey, and measures the four QLF parameters at
$z \sim 3.2$ and $z \sim 4$. M13 uses quasars identified in SDSS data in Stripe
82 (S82), and reports the four QLF parameters at $z \sim 5$. For all three
results, the parameters themselves and their associated 1$\sigma$ uncertainties
are reported. In the M13 results, the authors actually provide three different
fits to the observed results. In their fiducial result, they fix the value of
$\beta$, and fit for the three parameters $\log_{10}\phi^*$, $M^*$, and
$\alpha$. In a second set of parameters, the authors fix the value of $\alpha$
and find the best-fit values for the other three quantities. Finally, the
authors fix $M^*$, $\alpha$, and $\beta$, and only fit for
$\log_{10}\phi^*$. The best-fit values for the parameters change significantly
in some cases between the different fits. More importantly, none of these fits
seems to be ruled out conclusively by the data presented in M13, and so we
incorporate all of the fits in our results.

\begin{figure}[t]
  \centering
  \includegraphics[width=0.45\textwidth]{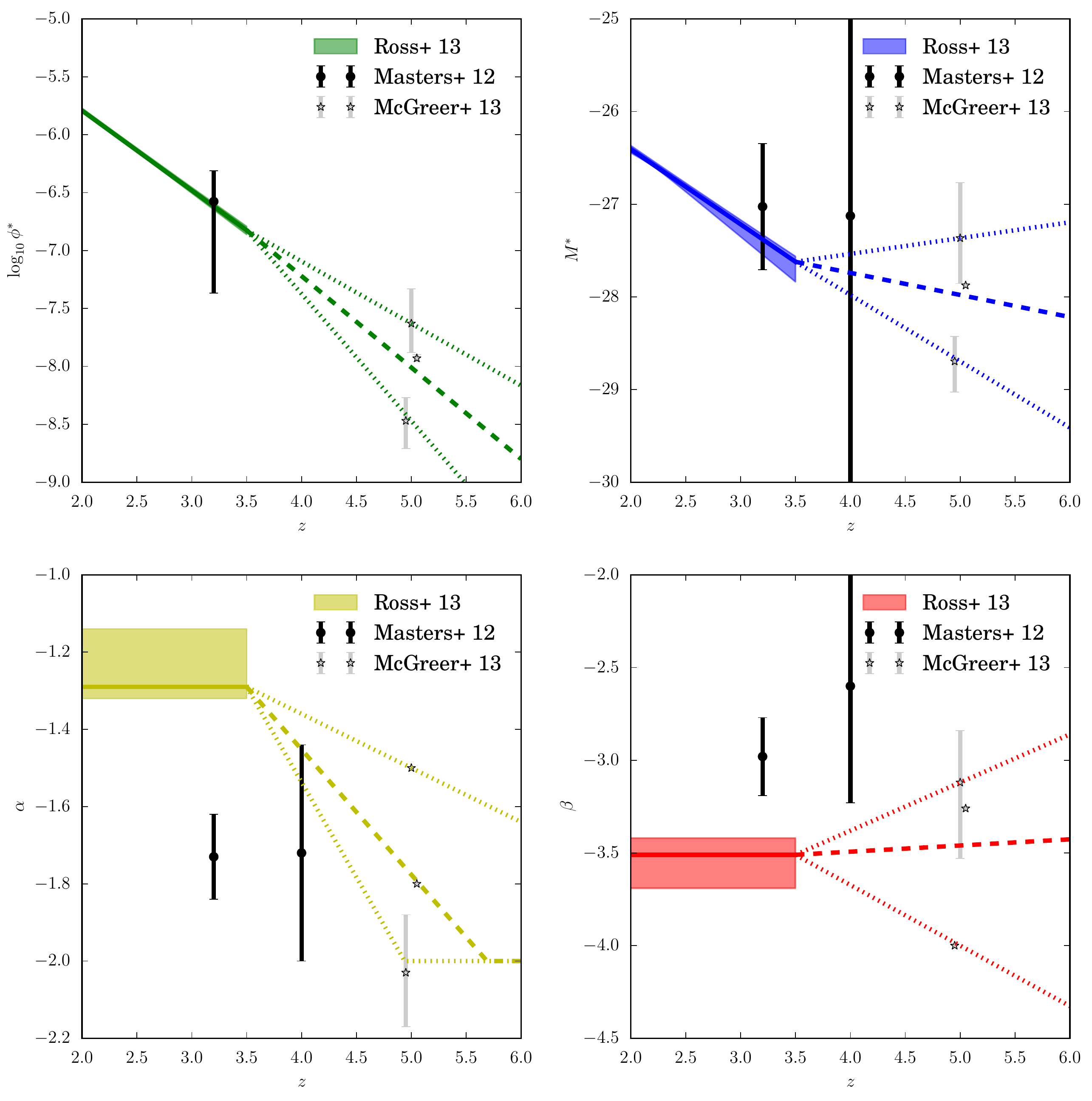}
  \caption{A plot of the evolution of the QLF parameters as a function of
    redshift: the base-ten logarithm of $\phi^*$ (top left), the break magnitude
    $M^*$ (top right), the faint-end slope $\alpha$ (bottom left), and the
    steep-end slope $\beta$ (bottom right). Best-fit values and associated
    1$\sigma$ errors from R13, M12, and M13 are represented as the solid lines
    with shaded error regions, dark-gray triangles, and light-gray stars,
    respectively. For the M13 data, all three sets of parameters provided by the
    authors are plotted at $z \sim 5$, slightly offset for visual clarity. The
    dashed lines for $z > 3.5$ show the fiducial evolution of the QLF, and the
    dotted lines show the bracketing ranges of values explored. See the text in
    this appendix for further details.}
  \label{fig:parameters}
\end{figure}

As explained in Sec.~\ref{sec:qlf}, our goal is to combine the observational
data from different epochs. For redshifts $z \leq 3.5$, the parameters from R13
are used. At higher redshift, the parameters are assumed to vary linearly in
redshift. The equations for the parameters are given in
Eqns.~(\ref{eqn:logphiz}-\ref{eqn:betaz}). The constant values are taken to be
those of R13 at $z=3.5$, and the slope of the redshift evolution is allowed to
take on a range of values. We will now discuss each of the four parameters in
turn.

For the parameter $\log_{10}\phi^*$, the fiducial value for the slope $c_1$ is
chosen to reproduce the average of the three reported values of M13 at
$z \sim 5$. As discussed in M13, the fits from R13 extrapolated to $z\sim 5$ do
not reproduce the overall normalization well, and predict too high a number
density. Thus, a steeper value than that of R13 is necessary. The range of
values for $c_1$ are chosen to bracket the range of best-fit values reported by
M13.

For the parameter $M^*$, the fiducial value of the slope $c_2$ is chosen to
reproduce the average of the three reported values of M13 at $z \sim 5$. The
slope is allowed to take on a range of values that bracket the three reported
values of M13. Also note that we have converted between magnitude systems using
$M_i(z=2) = M_{1450} - 1.486$, which assumes a spectral index $\alpha = 0.5$. If
instead the value of $\alpha = 0.61$ is used, as suggested by
\citet{lusso_etal2015} and used in the calculations of Sec.~\ref{sec:reion},
then conversion is $M_i(z=2) = M_{1450} - 1.681$. Further, if the SED from
\citet{shang_etal2011} is used, the conversion is $M_i(z=2) = M_{1450} - 2.139$.
The reason for the differences is that the $K$-corrections depend on the
spectral index of the SED (see Eqn.~3 of \citealt{richards_etal2006}). By
extension, the QLF can be affected when combining different data sets. However,
to be consistent with previous works that have combined disparate data sets in
this manner (\textit{e.g.}, R13 and M13), we use the conversion given by
assuming $\alpha = 0.5$.

For the parameter $\alpha$, the fiducial value of the slope $c_3$ is chosen to
reproduce the average of the three reported values of M13 at $z \sim 5$. As with
the other parameters discussed, a range of values is also explored which
brackets all of the reported values of M13. Further, the value of $\alpha$ is
bounded to lie where $\alpha > -2$. For $\alpha \leq -2$, the QLF does not
converge for low-luminosity objects, and a cutoff luminosity must be specified
below which quasars do not contribute significantly to helium reionization. To
avoid defining such a cutoff luminosity, the value of $\alpha$ is bounded. As a
practical matter, the ultimate goal of this project is to study helium
reionization using full numerical simulations, where the minimum resolved halo
mass will set the lower-limit of quasar luminosities.

Finally, for the parameter $\beta$, the fiducial value for the slope $c_4$ is
chosen to reproduce the average of the values from M13 at $z \sim 5$. The range
of slopes is chosen to bracket the values reported by M13. For the fiducial
choice of slope, the value of $\beta$ does not vary significantly with
redshift. This range of values incorporates much of the parameter space
constrained by M13, without the values of $\beta$ becoming arbitrarily
steep. However, the choice of $\beta$ ultimately does not significantly affect
the ionization level predicted by Eqn.~(\ref{eqn:volfrac}).

As a final note, the values of $\alpha$ and $\beta$ at $z \sim 3.2$ from M12 are
nominally inconsistent with the combined results from R13. However, when looking
at the results for individual redshift bins at $z \sim 3.2$ (\textit{e.g.},
Fig. 15 from R13), the uncertainties for the R13 values are significantly
larger, and the results are largely consistent at 1$\sigma$. The values of
$\log_{10}\phi^*$ and $M^*$ from M12 at $z \sim 3.2$ are consistent with the
results from R13, and those at $z \sim 4$ are consistent with the linear
redshift evolution given by the requirement of matching the M13 data. Note that
in Fig.~\ref{fig:parameters}, we do not plot the value of $\log_{10}\phi^*$ at
$z \sim 4$ from M12, because the reported lower-bound of the error bars is
larger than the best-fit value, which must be positive. Despite this fact, the
best-fit value is very close to the fiducial linear evolution given here.

Figure~\ref{fig:parameters} shows the measured parameters as a function of
redshift, as well as the assumed high-$z$ evolution for each parameter. The
solid lines and shaded regions show the best-fit parameters from R13, and the
individual points with error bars show the results from M12 and M13. The dashed
lines show the fiducial choices for the parameters, which are chosen as outlined
above. The dotted lines show the full range of parameters explored. The range of
parameter combinations is applied to helium reionization in
Figure~\ref{fig:volfrac} in the top-left panel. Note that, as discussed in
Sec.~\ref{sec:reion}, this uncertainty primarily affects the early stages of
reionization. Due to the recombination term in the calculation of the
volume-filling fraction and the fact that all reionization histories use the
parameters of R13 at $z \leq 3.5$, the high-$z$ values for the QLF do not
ultimately affect the timing of reionization significantly; nevertheless, the
different reionization scenarios can leave unique observable signatures on the
IGM.

\section{Bias as a function of redshift}
\label{sec:appendixb}
\begin{figure*}[t]
  \centering
  \resizebox{0.5\hsize}{!}{\includegraphics{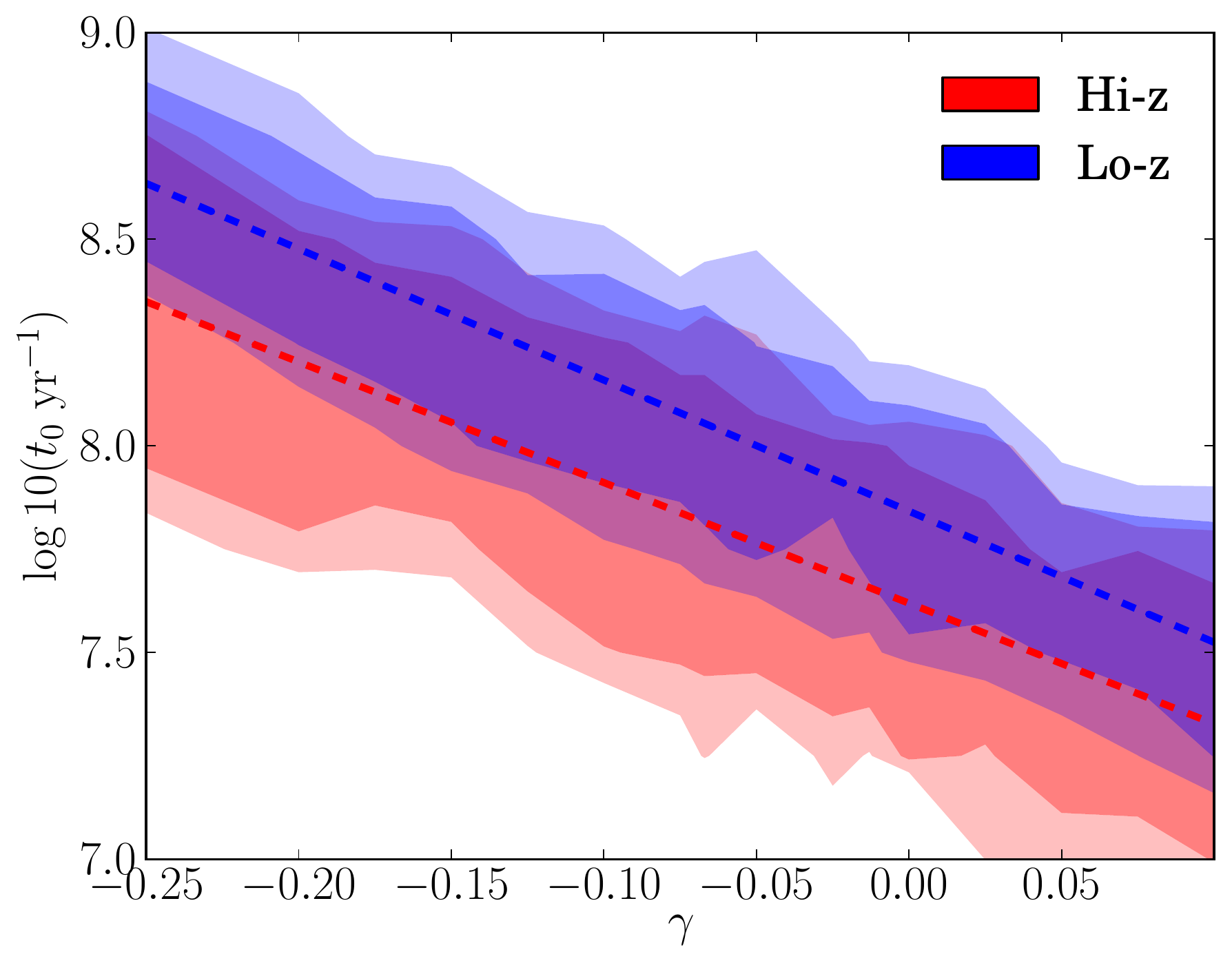}}%
  \resizebox{0.5\hsize}{!}{\includegraphics{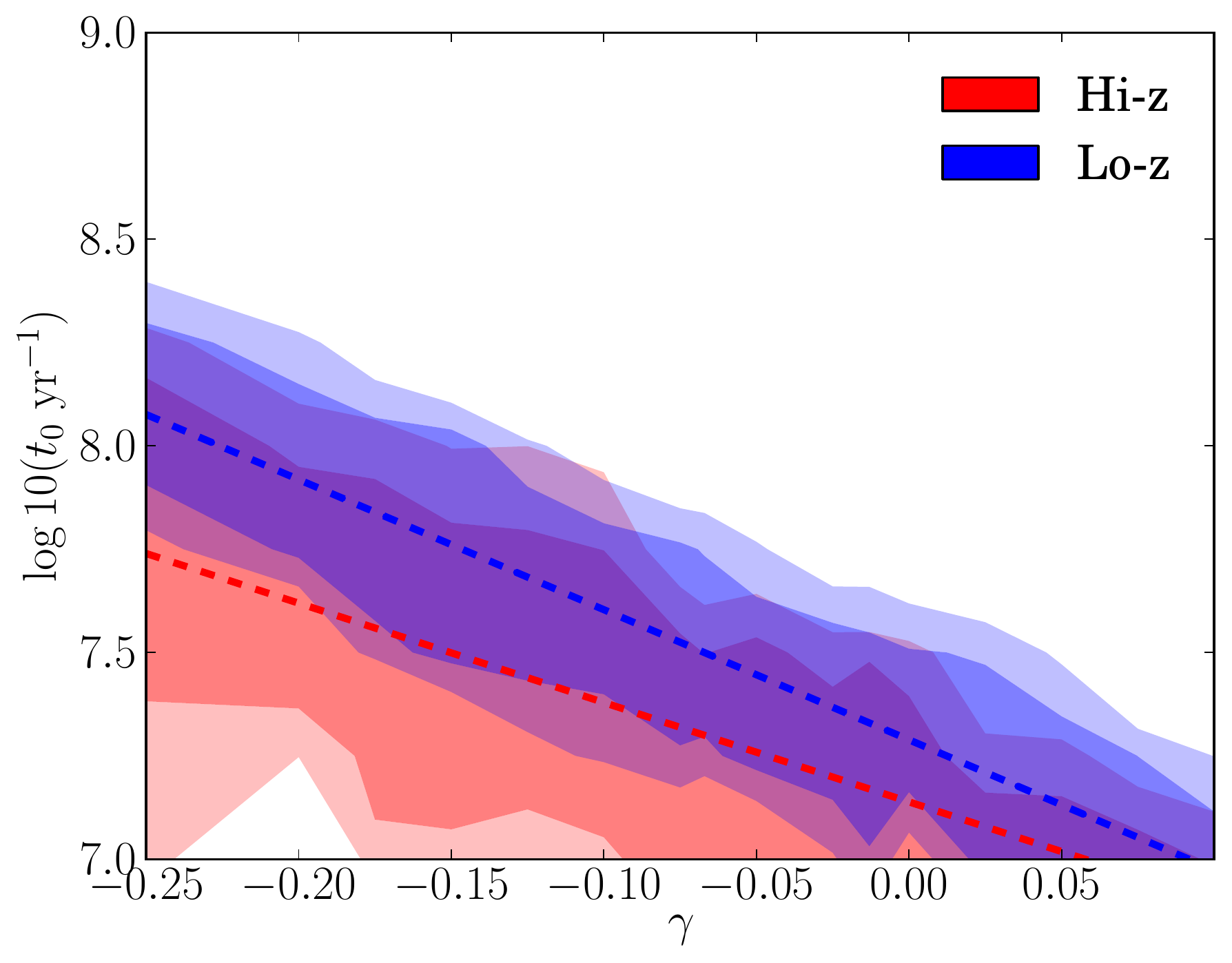}} \\
  \caption{Parameter space evolution of $t_0$ and $\gamma$ from
    Eqn.~(\ref{eqn:tq}) as a function of redshift for the lightbulb model (left)
    and the exponential model (right). As redshift decreases, the space of
    preferred models shifts slightly toward those with higher intrinsic
    clustering. This is in addition to the passive evolution in clustering
    signal that each individual model experiences, which constrains the space of
    applied models somewhat. Nevertheless, the results are consistent with there
    being no redshift evolution.}
  \label{fig:pspace-z}
\end{figure*}

In addition to reproducing the ``fiducial'' sample from the BOSS results, the
quasars from the constructed catalogs were also partitioned by redshift into a
``high-redshift'' and ``low-redshift'' sample in an analogous manner to the
auxiliary BOSS samples. In the case of the BOSS results, the ``fiducial'' sample
is actually the combination of the ``high-redshift'' and ``low-redshift''
samples, so these two datasets are statistically independent of each other, but
not the fiducial sample. For the purposes of comparing with the quasar catalogs,
however, it is possible to compute $\xi(s)$ at distinct points in redshift, and
compare with the BOSS results. The central redshifts for the high-redshift and
low-redshift samples are $z=2.51$ and 2.28, respectively. Then an analysis
similar to the above is performed, but at these additional redshifts. This
procedure yields further constraints on the bias as a function of redshift in
terms of the model parameters $t_0$ and $\gamma$. Figure~\ref{fig:pspace-z} is
similar to Figure~\ref{fig:pspace}, and shows how the selection of models varies
as a function of redshift. In general, we find that the choice of parameters for
our model $t_0$ and $\gamma$ evolves slightly with redshift. In general, the
BOSS measurements show an increase in bias with decreasing redshift. In order to
accommodate this increased bias, the model parameters must vary slightly. In
general, the model favors quasars with increased lifetimes as redshift
decreases. Despite this evolution with redshift, the relationship between
$\log_{10}(t_0)$ and $\gamma$ remains fairly linear, and it is still possible to
parameterize these models in terms of the characteristic lifetime and luminosity
factors $t_\mathrm{eff}$ and $L_\mathrm{eff}$ as defined in
Eqn.~(\ref{eqn:teff}).

Table~\ref{table:teff} summarizes the changes in best-fit parameters as a
function of redshift. Interestingly, these values change somewhat: as structure
continues to build, models with increasingly higher bias values are
preferred. The fact that the best-fit values change demonstrates that the
passive evolution of an increased clustering signal within a given model is not
sufficient; rather, this redshift evolution introduces additional constraints
that we can use to select the most appropriate model. Nevertheless, the results
are consistent with no redshift evolution. The results of \citet{white_etal2012}
also suggest that redshift evolution is minimal. Extending the clustering
measurements to a larger redshift range could provide important constraints on
the properties of quasar hosts.

\section{Bias as a function of luminosity}
\label{sec:appendixc}

\begin{figure}[t]
  \centering
  \includegraphics[width=0.45\textwidth]{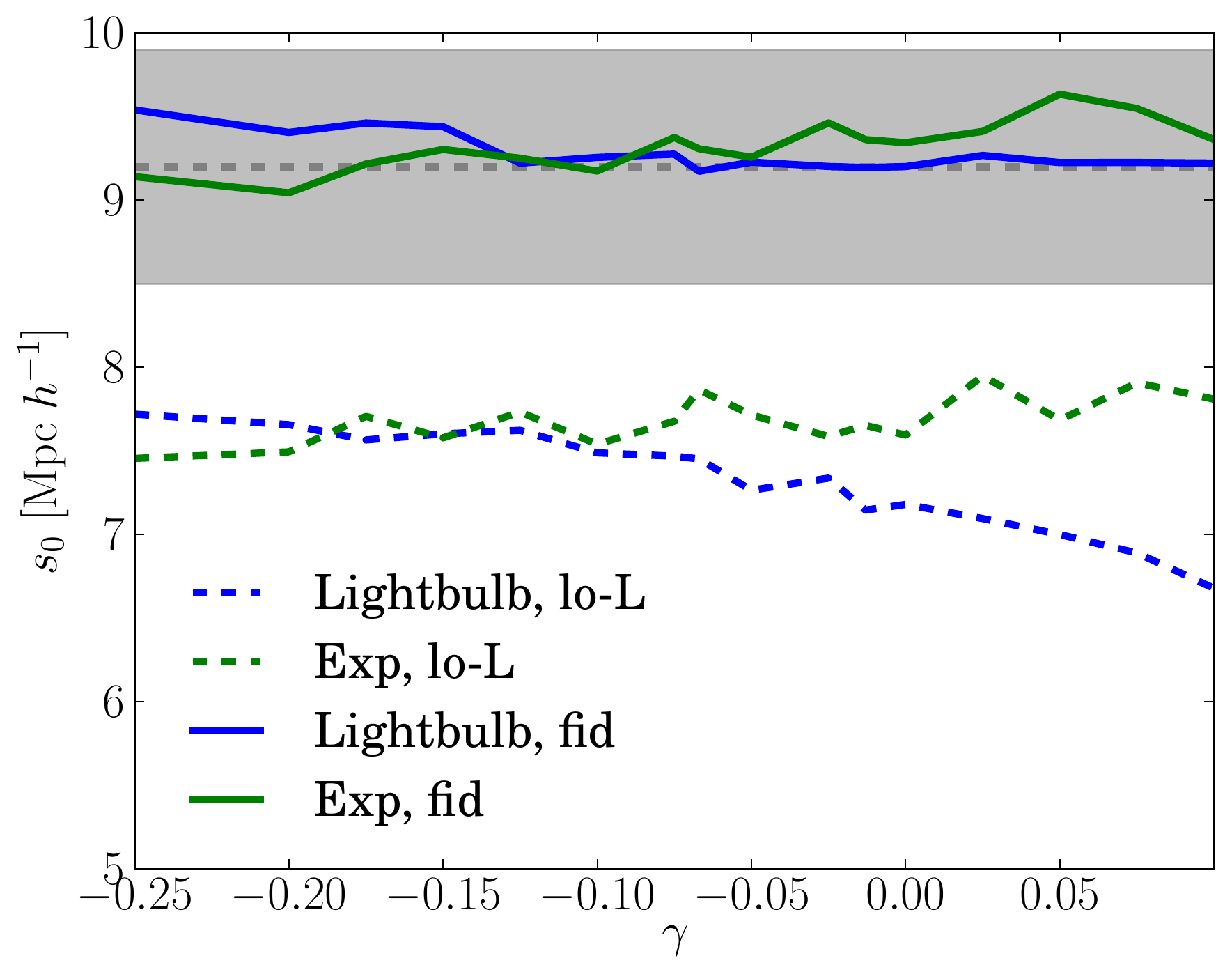}
  \caption{The best-fit parameter $s_0$ for the two-point correlation function
    in the form $\xi(s) = \qty(s/s_0)^{-2}$ as a function of power-law index
    $\gamma$ from Eqn.~(\ref{eqn:tq}) for the lightbulb and exponential cases
    using a fiducial (solid) and low-luminosity (dashed) luminosity
    selection. The gray shaded region shows the BOSS measurement for the
    fiducial luminosity cut. For the low-luminosity quasars, we see opposite
    trends for the two models. For the lightbulb, more negative values of
    $\gamma$ mean that dimmer quasars have longer lifetimes, which combined with
    abundance matching implies they have more massive hosts. They therefore have
    larger values of $s_0$ compared to more positive values of $\gamma$. In the
    exponential case, larger values of $\gamma$ show more clustering because the
    bright quasars are longer lived, and are more likely to be included in the
    low-luminosity cuts while they are below their peak luminosity. Since they
    are abundance matched to more massive, highly clustered hosts, this leads to
    the behavior seen. See the text for further discussion.}
  \label{fig:lum}
\end{figure}

We can also examine the dependence of bias as a function of quasar
luminosity. In the preceding analysis, we looked at the fiducial luminosity
selection of the BOSS measurements for clustering, $-25 \geq M_i \geq -27$. In
order to break the degeneracy in Fig.~\ref{fig:pspace}, we explored the
implications of measuring the clustering of quasars with different luminosity
cuts. We examined a high-luminosity cut $M_i \leq -27$, and a low-luminosity cut
$-23 \geq M_i \geq -25$. Unfortunately, since the simulation volumes are only
1~($h^{-1}$~Gpc)$^3$, there are an insufficient number ($\sim$400) of
high-luminosity objects to constrain the two-point correlation function.

When fitting the functional form of the two-point correlation function, a power
law is used:
\begin{equation}
\xi(s) = \qty(\frac{s}{s_0})^\beta.
\label{eqn:xi}
\end{equation}
Fits the function are made for cases where the exponent $\beta$ is allowed to
vary, and others with a fixed value of $\beta=-2$ as in
\citet{white_etal2012}. In both cases, the clustering length $s_0$ increases for
larger values of the bias. To fit the best parameters, the parameters $s_0$ and
$\beta$ that minimized the $\chi^2 = \delta^\mathrm{T} C^{-1} \delta$ value were
found, where $\delta$ is defined as the difference between the average $\xi(s)$
and the functional form and $C$ is the covariance matrix, calculated in the same
way as in Sec.~\ref{sec:chi2}. These fits were made for the best-fit models
defined in Eqn.~(\ref{eqn:teff}) using the values in Table~\ref{table:teff}.

Figure~\ref{fig:lum} shows the value of the correlation length fits $s_0$ for
the fiducial luminosity cut $-25 \geq M_i \geq -27$ (solid lines) and the
low-luminosity cut $-23 \geq M_i \geq -25$ (dashed lines) for the lightbulb and
exponential models. The data are somewhat noisy, owing to the comparatively
large shot-noise error in the correlation function measurement. However, there
does seem to be a trend emerging: in the lightbulb case, for more negative
values of $\gamma$, the bias is larger, with the opposite trend for the
exponential case. In the lightbulb case, this can be explained by noting, as in
Sec.~\ref{sec:chi2}, that in abundance matching longer lifetimes lead to a
larger bias in the host halos. For negative values of $\gamma$, less luminous
quasars have longer lifetimes. Subsequently these quasars are being hosted in
more massive halos. This means the clustering is stronger for large negative
values of $\gamma$, implying a larger value of $s_0$.

In the exponential case, the opposite trend is observed due to the presence of
high-$L_\mathrm{peak}$ interlopers. For positive values of $\gamma$, brighter
quasars have longer lifetimes, and are more likely to be included in the
low-luminosity selection. Since these hosts are abundance matched to occupy more
massive, more clustered halo hosts, this leads to a stronger clustering signal,
and a larger value of $s_0$. The evolution is not as strong as in the lightbulb
case, however. In principle, the clustering measurement in different luminosity
ranges could help break the degeneracy of best-fit models.

Unfortunately, in practice this type of measurement might be difficult to
actually make. The change in bias between the extreme values of $\gamma$ is not
very significant, and the measurement is very noisy. The shaded gray region in
Figure~\ref{fig:lum} shows the current 1$\sigma$ bounds from the BOSS
measurement, which has a larger spread than the variation in $s_0$ as a function
of $\gamma$. Nevertheless, this ratio is a possible way to break the degeneracy
between the different models.
\end{appendix}

\end{document}